\documentclass[%
 aapm,
 mph,%
 amsmath,amssymb,
 reprint%
]{revtex4-2}

\usepackage{graphicx}
\usepackage{dcolumn}
\usepackage{bm}

\usepackage[mathlines]{lineno}

\newcommand{\fm}{\mbox{fm}}

\newcommand{\MeV}{\mbox{MeV}}
\newcommand{\GeV}{\mbox{GeV}}

\newcommand{\plaq}{\mbox{pl}}
\newcommand{\rt}{\mbox{rt}}
\newcommand{\Tr}{\mbox{Tr}}
\newcommand{\sign}{\mbox{sign}}
\newcommand{\spec}{\mbox{spec}}
\newcommand{\imp}{\mbox{imp}}
\newcommand{\ov}{\mbox{ov}}

\begin{document}

\preprint{AAPM/123-QED}

\title[]{Vector $K^{*}$ mesons in external magnetic field from  SU(3) gluodynamics}

\author{E.V. Luschevskaya}
 \email{luschevskaya@itep.ru, luschevskaya@gmail.com}
\affiliation{National Research Center “Kurchatov Institute”, Moscow, 123182 Russia,}

\affiliation{Moscow Institute of Physics and Technology, Dolgoprudnyj, Institutskij lane 9, Moscow Region 141700, Russia}

\author{O.V. Teryaev}%
\email{teryaev@theor.jinr.ru}
\affiliation{Joint Institute for Nuclear Research, Dubna, 141980, Russia}

\affiliation{National Research Center “Kurchatov Institute”, Moscow, 123182 Russia}

\author{E.A. Dorenskaya}
 \email{dorenskaya@itep.ru}
\affiliation{National Research Center “Kurchatov Institute”, Moscow, 123182 Russia}

\date{\today}

\begin{abstract}
We explore the ground state energies of  the neutral and charged  vector $K^{*}$ mesons in the external abelian  magnetic field of QCD scale using overlap fermions. We attempt to calculate the magnetic moment and  the  magnetic dipole polarizability  of the $K^{*0}$ and $\bar{K}^{*0}$ mesons and   investigate  their dependence  on the ratio of the strange  quark mass to the light quark mass. It is found that the ground state energy and the  magnetic dipole polarizability depends on the meson spin projection on the magnetic field axis. This leads  to the appearance of dileptonic asymmetry, which can be characterized by the tensor polarizability   estimated for the $K^{*0}$ mesons. We obtain that the $g$-factor   of   the vector $K^{*\pm}$ mesons   depends on  the   $m_s/m_u$ value.  Also we do not observe any evidence of the tachyon mode existence for the strange vector mesons.
\end{abstract}

\maketitle

\section{Introduction}{\label{intro}}
Quantum chromodynamics in strong magnetic fields is a rich realm of research for theoretical and experimental physics.
  The strong magnetic fields of  QCD scale arise  in  non-central heavy-ion collisions at LHC, SPS,  AGS, RHIC in terrestrial laboratories such as CERN, BNL  and  in future experiments  FAIR and NICA  in  GSI  and JINR. One of the most important tasks of these experiments is an exploration of a quark-hadronic matter under extreme conditions. It is found that  the lowest bound on the magnetic field value at the LHC (CERN) parameters has to be equal $eB\sim 15\times m^2_{\pi}$ and $eB \sim m^2_{\pi}$ for the RHIC energies (BNL) \cite{Toneev}.

   Such strong magnetic fields lead to a variaty of remarkable phenomena, among them the chiral magnetic effect \cite{Kharzeev:parity, Kharzeev:topQ}, the inverse magnetic catalysis \cite{Bruckmann:catalysis},  the modification of QCD phase diagram, in particular the decrease  of the chiral symmetry breaking temperature and the confinement-deconfinement transition temperature  with the magnetic field growth  \cite{Bali:QCDPhD,Endrodi:QCDPhD,Ilgenfritz:MagCatalysis,Bornyakov:DecTransition}, the   dilepton emission asymmetry  \cite{Terayev}.
  Lattice QCD simulations and   phenomenological models show that  the masses of   hadrons depend on the magnitude of the magnetic field.   Therefore, it may  affect hadronization processes in heavy-ion collisions.  Thus,  the mass of the neutral pion $\pi^0$ decreases, while the masses of charged pions $\pi^{\pm}$ increase with the growth of the magnetic field strength \cite{Simonov:2013a,Luschevskaya:2015a,Bali:2015,Luschevskaya:2015b,Luschevskaya:2016,Bali:2018}.

Quite contradictory results have been obtained  for  vector $\rho$ mesons in external magnetic field. In a sufficiently strong magnetic field  the QCD vacuum becomes a superconductor \cite{Chernodub:2010,Chernodub:2011,Braguta:2012,Chernodub:2014,Liu:2015}. This phenomenon is accompanied by the condensation of the vector $\rho^{\pm}$ mesons. However,  there are lattice calculations and  investigations within the framework of phenomenological models, which reveal the absence of the tachyonic mode \cite{Simonov:2013a, Andreichikov:2016,Luschevskaya:2015a,Bali:2015,Luschevskaya:2015b,Bali:2018, Luschevskaya:2013,Hidaka:2013,Taya:2015,Luschevskaya:2017}. 
This does not mean the absence of phase transitions, and such a transition was recently observed and explored in the lattice studies of electroweak theory \cite{Chernodub:2022ywg}.

In QCD, masses of the $\rho^{\pm}$   mesons are varying smoothly and do not go to zero due to their nonlinear response in an external strong  magnetic field. The contribution of these nonlinear effects can be assessed using  the magnetic dipole polarizability and hyperpolarizabilities.
 The magnetic  polarizabilities of hadrons can be explored not    only  by the theoretical methods of QCD \cite{Luschevskaya:2016,baldin,Lee:2005,Gasser:2005,Aleksejevs:2013,Savage:2015}, but also can be  measured in experiments \cite{Antipov:1983,Filkov:2006,Adolph:2015}. The energy spectrum of charmonium and $D$ mesons has been  studied in a strong  magnetic field within the framework of QCD sum rules  \cite{Cho:2015,Gubler:2015}.

We continue to  explore the  energy of the neutral and charged vector mesons  considering  the strange $K^*$ mesons. We check whether their energies might turn to zero at large fields, but do not find any hint of such behavior.
 The vector  $K^*$ mesons also posses nonzero magnetic dipole polarizability and hyperpolarizabilities which lead to the nonlinear energy response to an external strong  magnetic field.  Already at the magnetic fields $eB\lesssim 1.2\ \GeV^2$ the energy of $K^{*\pm}$ mesons tends to form plato for the low energy branch preventing the tachyonic mode formation.

The magnetic moment of meson is another important quantity that characterizes the linear response of the  charged meson to the external magnetic field.  The magnetic moments of mesons and baryons have been calculated in lattice QCD
 \cite{Luschevskaya:2017,Martinelli:1982,Andersen:1996,Lee:2005baryon,Hedditch:2007,Lee:2008,Owen:2015,Savage:2017,Parreno:2017}, in  low-energy effective field theory of strong interactions \cite{Djukanovic:2013}, using the covariant quark model \cite{Melo:1997},  the light cone quark model \cite{Choi:2004}, the Dyson-Schwinger equations of QCD \cite{Bhagwat:2008}, the QCD sum rules \cite{Samsonov:2003,Aliev:2009} and the field cumulant  method \cite{Simonov:2013b}.
It is known that the  neutral   $K^{*0}$ and $\bar{K}^{*0}$ mesons have nonzero magnetic moment   \cite{Hedditch:2007,Lee:2008,Bhagwat:2008,Aliev:2009}.      However, there is no agreement on the sign and magnitude of the magnetic moment obtained in different approaches. For example the lattice calculations   \cite{Hedditch:2007} gives the value $g=-0.26$    for the $K^{*0}$ meson against the  $g = 0.26\pm0.4$ obtained in the framework of light cone QCD sum rules \cite{Aliev:2009}. The value $g(K^{+*})=2.23$  was obtained from the lattice calculations using 3-point correlation function method \cite{Hedditch:2007}. The lattice background field approach gave the value $g(K^{+*})=2.36$. This approach uses the 2-point correlation functions  for the calculation of a meson energy shift \cite{Lee:2008}.
The light-cone QCD sum rules   predicted   $g=2.0\pm0.4$ \cite{Aliev:2009}. The  value $g=2.08$ was found from the Dyson-Schwinger equations of QCD \cite{Bhagwat:2008} and $g=2.194$ from the field cumulant  method \cite{Simonov:2013b}.  

 \section{Technical details of the simulations}
 \label{sec-2}
 
 \ \ \ \ \ We have carried out  our calculations  in $SU(3)$  pure lattice gauge theory in an external abelian strong magnetic field.
 For the generation of statistically independent ensembles of gauge field configurations we use the improved L\"uscher-Weisz action \cite{Luscher:1985}
 
\begin{equation}
S=\beta_{\imp} \sum_{\plaq} S_{\plaq}-\frac{\beta_{\imp}}{20 u^2_0}\sum_{\rt}S_{\rt}.
\end{equation}

The  plaquette and rectangular loop terms are represented by  $S_{\plaq,\rt}=(1/3)\Tr(1-U_{\plaq,\rt})$, where $u_0 = (\langle (1/2) {\rm Tr} U_{\plaq} \rangle)^{1/4} $ is the input tadpole factor   computed at zero temperature \cite{Bornyakov:2005}. We consider $150-200$ lattice gauge field configurations for   the lattice volume $N_s^3 \times N_t=18^4$ and the lattice spacing $a=0.105\ \fm$  which corresponds to  the action parameter  $ \beta_{\imp} = 8.30$.

To obtain the $K^*$ meson correlation  functions   we  calculate  the quark  propagators which can be approximated by the following series

\begin{equation}
D^{-1}(x,y)=\sum_{k<M}\frac{\psi_k(x) \psi^{\dagger}_k(y)}{i \lambda_k+m_q},
\label{lattice:propagator}
\end{equation}

where $\psi_k(x)$ is the $k$-th  eigenfunction, $\lambda_k$ is the  eigenvalue corresponding to the $k$-th  eigenfunction, $M=50$  is the number of eigenmodes  sufficient for the approximation of the proppagator, $m_q$ is the bare quark mass of the light  or strange quark.

To obtain the $M$ lowest eigenfunctions and eigenvalues    we solve numerically the Dirac equation

\begin{equation}
D \psi_k=i \lambda_k \psi_k, \  \ D=\gamma^{\mu} (\partial_{\mu}-iA_{\mu}).
\label{Dirac}
\end{equation}

The total gauge field $A_{\mu}$ is the sum of the  $SU(3)$ gauge field  of gluons $A_{\mu}^{gl} = g t^a A^a_{\mu} $ and external $U(1)$  field $A_{\mu}^{B}$

 \begin{equation}
A_{\mu}= A_{\mu}^{gl} + A_{\mu}^{B},
\label{gaugefield}
\end{equation}

where $\mu = 0,1,2,3$ is the Lorentz index, $g$ is the strong interaction constant, $t^a$ are the generators of the $SU(N_c)$ group and $a = 1, ... ,N^2_c -1 $ is the color index.
 The  external magnetic field $B$  is directed  along the $z$-axis
 
\begin{equation}
 A^B_{\mu}(x)=\frac{B}{2} (x_1 \delta_{\mu,2}-x_2\delta_{\mu,1}).
\end{equation}

 It is quantized on the lattice according to the relation  \cite{Al-Hashimi:2009,Hooft:1979,Zainuddin:1989,Chen:1996}

 \begin{equation}
qB = \frac{2 \pi k} {(aN_s) ^ 2}, \ \ k \in \mathbb {Z},
\label{quantization}
\end{equation}

where $N_s$ is the number of lattice sites in the spatial dimension, $a$ is the lattice spacing,  $ q = - 1/3 \, e $ is the quark charge.
This quantization condition arises because the gauge invariance must not be violated and periodic boundary conditions in space for fermions have to be satisfied. The  transformations

 \begin{widetext}
\begin{eqnarray}
\psi(x_1+aN_s,x_2,x_3) = \exp(-i\frac{q}{2}B a N_s x_2)\psi(x_1,x_2,x_3),\ \ \ \ \psi(x_1,x_2+a N_s,x_3)=\exp(i\frac{q}{2}B aN_s x_1)\psi(x_1,x_2,x_3)
\label{twbc1}
\end{eqnarray}
\end{widetext}

satisfy both these requirements and give the condition \eqref{quantization}.

The  magnetic field \eqref{quantization} is taken into account  only in  Dirac operator, because the abelian fields interact  only with quarks and we do not consider the dynamical quark loops in our   simulations. To calculate the Dirac spectrum we use the  Neuberger overlap operator  \cite{Neuberger:1997} preserving chiral invariance on the lattice at zero quark mass. It   has the following form

 \begin{equation}
 M_{\ov}=\left(1-\frac{am_q}{2 \rho}\right) D_{\ov}+m_q,
 \end{equation}
 
where $m_q$ is the bare quark mass, $D_{\ov}$ is the massless overlap operator defined by

\begin{equation}
 D_{\ov}= \frac{\rho}{a} \left( 1+
  \frac{D_W}{\sqrt{D^{\dagger}_W D_W}} \right). \label{overlap}
\end{equation}

 In formula \eqref{overlap}  $D_W=M-\rho/a$ is the Wilson-Dirac operator,  $M$ is the Wilson hopping term with $r=1$, $\rho=1.4$ is the parameter.

The hermitian Wison-Dirac operator $H=\gamma_5 D_W $ determines  the $\sign(H)$ function

 \begin{equation}
\sign(H)=\frac{H}{ \sqrt{H^{\dagger} H}}= \frac{H}{||H||} = W,
\label{sign_function}
\end{equation}

then the operator \eqref{overlap}  can be rewriten in the folowing way

\begin{equation}
D_{\ov}= \frac{\rho}{a} \left( 1+\gamma_5\sign(H) \right).
\label{overlap_}
\end{equation}

The  Dirac spectrum were calculated at the bare  light quark mass $m_q a=0.007$ corresponding to the pion mass $m_{\pi}=367(8)\,\MeV$ and the  bare   strange quark masses  $m_s a = 0.02$,\,$0.04$,\,$0.06$,\,$0.08$,\,$0.1$,\,$0.14$,\,$0.175$,\,$0.21$. Thus, we     explore the dependence of the dipole polarizability $\beta_m$ and $g$-factor on the $m_s/m_{u,d}$ ratio.

For the hermitian Wilson-Dirac operator $\spec(H) \in [\lambda_{\min},\lambda_{\max}]\in {\cal R}$, also $||H||=\lambda_{\max}$, therefore $\spec(W)\in
[\lambda_{\min}/\lambda_{\max};1]$. The   linear combinations

\begin{equation}
P_n(H^2)=\sum_{k=0}^{n}c_k T_k(z),\ \ z=\frac{2H^2-1-\epsilon}{1-\epsilon},
\end{equation}

of  the Chebyshev polynomials  $T_k(z)$, $k=0,..,n$ approximate the sign function \eqref{sign_function} at $\sqrt{\epsilon}\leqslant \spec(H)
\leqslant 1$, where $\epsilon=\lambda^2_{min}/\lambda^2_{max}$ \cite{Giusti:2003}. Therefore, the polynomials $P_n(H^2)$ and  the  $H$ matrix
  have the same  set of eigenfunctions $\psi_k$    \cite{Neff:2001}.
  
 \section{Meson correlation functions}
  \label{sec-3}
   The correlation functions of the $K^{*0}$ and $\bar{K}^{*0}$ mesons  have the following  form
  \begin{equation}
 \langle \bar{\psi}_{d,s}(x) \gamma_i \psi_{s,d}(x) \bar{\psi}_{s,d}(y) \gamma_j \psi_{d,s}(y) \rangle_A.
 \label{observables_Kq0}
 \end{equation}
For the charged $K^{*\pm}$ mesons they we exploit the following correlator
\begin{equation}
 \langle \bar{\psi}_{u,s}(x) \gamma_i \psi_{s,u}(x) \bar{\psi}_{s,u}(y) \gamma_j \psi_{u,s}(y) \rangle_A,
 \label{observables_Kq1}
  \end{equation}
   where we consider   the Dirac gamma matrices $ \gamma_i, \gamma_j $ only with the  spatial indices $i,j = 1,2,3$ and     $\psi^{\dagger} = \bar{\psi}$ in the Euclidean space. The lattice coordinates
$ x = (\textbf {n} a, n_ta) $ and $ y = (\textbf{n}^{\prime} a, n^{\prime}_t a) $ specify the quark positions  in the lattice volume,
where $ \textbf{n}, \textbf{n}^{\prime} \in \Lambda_3 = \{(n_1, n_2, n_3) | n_i = 0,1, ..., N_s -1 \}$, $n_t, n^{\prime}_t = 0,1,...,N_t-1$.

 To calculate the correlation functions \eqref{observables_Kq0} and \eqref{observables_Kq1} we use the
  equalities
\begin{widetext}
\begin{eqnarray}
\langle \bar{\psi}_{d,s}(x) \gamma_i \psi_{s,d}(x) \bar{\psi}_{s,d}(y) \gamma_j \psi_{d,s}(y) \rangle_A=-\Tr[\gamma_iD^{-1}_{s,d}(x,y)\gamma_jD^{-1}_{d,s}(y,x)],\label{lattice:correlatorKq0}\\
\langle \bar{\psi}_{s,u}(x) \gamma_i \psi_{u,s}(x) \bar{\psi}_{u,s}(y) \gamma_j \psi_{s,u}(y) \rangle_A=-\Tr[\gamma_iD^{-1}_{u,s}(x,y)\gamma_jD^{-1}_{s,u}(y,x)],
\label{lattice:correlatorKq1}
\end{eqnarray}
\end{widetext}
where
 the propagators  $D^{-1}_d$, $D^{-1}_u$ and $D^{-1}_s$  are calculated on the lattice with the use of the relation \eqref{lattice:propagator}.
At zero magnetic field $D^{-1}_d=D^{-1}_u$, but at $B\neq0$ the isospin symmetry is broken and we take into account the twice different charges of $d$ and $u$ quarks.
We perform the spatial Fourier transform  of  the right parts of the relations \eqref{lattice:correlatorKq0}
  and \eqref{lattice:correlatorKq1}  from the coordinate to the momentum space. Zero spatial momentum corresponds to the  ground state energy of  mesons.

  We use the covariant density matrix to define the correlation functions of the ground state energies of the vector $K^*$ mesons   in the external magnetic field.
  If the spin density matrix is expressed  in terms of the transverse ($e_x=(0,1,0,0), e_y=(0,0,1,0)$) and longitudinal ($e_z=(0,0,0,1)$) polarization vectors,
then      the energies   of the  $K^{*}$  mesons with   the spin projection $S_z=0$  are obtained from the correlation function
\begin{equation}
C(S_z=0)=\langle O_3 (t)\bar{O}_3 (0)\rangle_A,
\label{eq:CVV0}
    \end{equation}
    where
 $O_3=\psi^{\dagger}_{s}(x) \gamma_3 \psi_{d}(x)$    and  $O_3=\psi^{\dagger}_{d}(x) \gamma_3 \psi_{s}(x)$
 are the interpolation operators of the $K^{*0}$ and $\bar{K}^{*0}$ mesons respectively. The  operators $O_3=\psi^{\dagger}_{s}(x) \gamma_3 \psi_{u}(x)$ and $O_3=\psi^{\dagger}_{u}(x) \gamma_3 \psi_{s}(x)$ correspond  to the $K^{*+}$ and $K^{*-}$ mesons.

The energies of   vector mesons with the spin projections $S_z=-1$ and $S_z=+1$ are obtained from the combinations of correlation functions
\begin{eqnarray}
&&C(S_z=\pm 1)=\langle O_1 (t)\bar{O}_1 (0)\rangle_A+\langle O_2 (t)\bar{O}_2 (0)\rangle_A \nonumber\\
&& \pm i(\langle O_1 (t)\bar{O}_2 (0)\rangle_A-\langle O_2 (t)\bar{O}_1 (0)\rangle_A)
\label{eq:CVV1}
\end{eqnarray}
where
 $O_1=\psi^{\dagger}_{d,s}(x) \gamma_1 \psi_{s,d}(x),\,  O_2=\psi_{d,s}^{\dagger}(x) \gamma_2 \psi_{s,d}(x)$
correspond to the  $K^{*0}$ and $\bar{K}^{*0}$ mesons   and  $O_1=\psi^{\dagger}_{u,s}(x) \gamma_1 \psi_{s,u}(x),\,  O_2=\psi_{u,s}^{\dagger}(x) \gamma_2 \psi_{s,u}(x)$  are for the  $K^{*\pm}$ mesons.

One can expand the correlation function in a series over the eigenstates of the Hamiltonian $\widehat{H}$ of the theory
\begin{eqnarray}
&& \langle O_i(t)  \bar{O}_j(0)\rangle_T= \nonumber \\
&&=\frac{1}{Z}\sum_{m,n}e^{-(T-t)E_m}\langle m|\widehat{O}_i|n \rangle e^{-tE_n}\langle n| \widehat{O}_j^{\dagger}|m \rangle,
\label{series}
\end{eqnarray}

where $i,j=1,2,3$, $E_m$ ans $E_n$ are the energy  eigenvalues of $\widehat{H}$ and the partition function reads
\begin{equation}
 Z=\sum_n \langle n|e^{-T\widehat{H}} |n \rangle=\sum_n e^{-TE_n}.
 \label{Zfunc}
\end{equation}

In  thermodynamical  limit the expression  \eqref{series} takes the form
\begin{equation}
\langle O_i(t)  \bar{O}_j(0)\rangle_{T\rightarrow \infty}=  \sum_{n} \langle 0|\widehat{O}_i|n \rangle \langle n| \widehat{O}_j^{\dagger}|0 \rangle e^{-t E_n}.
\label{sum}
\end{equation}

The main contribution in \eqref{sum}  comes from the ground state energy    at large  $n_t = t/a$. Therefore, taking into account the  periodic boundary conditions in the leading order  we obtain   the relation

 \begin{equation}
\tilde{C}(n_t) = 2A_0 e^{-N_T a E_0/2} \cosh ((\frac{N_T}{2}-n_t) a E_0),
 \label{coshfit}
\end{equation}

where $A_0$ is some constant, $E_0$ is the energy of the ground state   \cite{Lang:2010}.

 To reduce statistical uncertainties at zero magnetic field  we   averaged the correlators $\tilde{C}(n_t)$  over the
 three  spatial directions.   At nonzero field we  performed the averaging   of the correlation functions over the opposite directions of the magnetic field if the corresponding data was available.  The method for calculating the effective mass from these averaged correlation functions does not differ from that presented in our previous work \cite{Teryaev:Tensorpolar}.

\section{Vector $K^{*0}$ and $\bar{K}^{*0}$ mesons}
 \label{sec-4}
 
\subsection{Energy of $d\bar{s}$ and $s\bar{d}$ states with $J=1$}
 \label{subsec-41}
 
 \ \ \ \ \ In an external   magnetic field the energy squared  of a point-like particle  is described by the following formula
\begin{equation}
E^2=p^2_z+(2n+1)|qB|-gS_zqB+m^2,
\label{eqLL}
\end{equation}
where $p_z$ is the momentum in the 'z' spatial direction, $n$ is the
principal quantum number, $q$ is the electric charge of the
particle, $g$ is the  g-factor, $S_z$ is the spin projection on the
field direction and $m\equiv E(B=0)$  is the energy of the particle
at zero magnetic field  and zero momentum.

If we consider the strongly interacting particle in a strong  magnetic field, then the meson energy  $\eqref{eqLL}$ gets the additional nonlinear corrections \cite{Luschevskaya:2015a,Luschevskaya:2017}.
There is a deviation from the expected   linear behaviour because the internal structure of  mesons is getting sensitive to the external magnetic field when it becomes equal or higher than the QCD scale. The contribution of these corrections can be characterized by the values of the  dipole polarizability  $\beta_m$ and hyperpolarizabilities of higher orders $\beta_m^{h1}$, $\beta_m^{h2}$, etc. 
The different masses of the constituent quarks in  $K^{*0}$ and $\bar{K}^{*0}$ mesons   leads to not  large, but nonzero magnetic moment, which we also include into consideration.

For the ground state energy of mesons at rest we take  $n=0$ and $p_z=0$.
  For the neutral   $K^{*0}$ and $\bar{K}^{*0}$ mesons   we  also put  $q=0$.
  In a relativistic case the energies squared    for $S_z=0$ and $S_z=\pm 1$ spin projections have the following form

\begin{widetext}
\begin{eqnarray}
&&E^2(S_z=0) = m^2 -4 \pi m \beta_m (eB)^2 - 4 \pi m \beta_m^{h1} (eB)^4 -  4 \pi m \beta_m^{h2}(eB)^6 - 4 \pi m \beta_m^{h3} (eB)^8-... ,
\label{eq:K0:s-0:B8}\\
&&E^2(S_z=\pm 1) = m^2 \mp g(eB)-4 \pi m \beta_m (eB)^2 - 4 \pi m \beta_m^{h1} (eB)^3 -4 \pi m \beta_m^{h2}(eB)^4 -  4 \pi m \beta_m^{h3} (eB)^5 - ...\,.
\label{eq:K0:s1:B5}
\end{eqnarray}
\end{widetext}
We  require the parity  conservation, so for the spin projection $S_z=0$ the energy squared has to include  only the terms of even powers of the magnetic field, see the formula \eqref{eq:K0:s-0:B8}. For the spin projections $S_z=+1$ and $S_z=-1$ the even and odd powers of the magnetic field are allowed \eqref{eq:K0:s1:B5}.

At relatively small magnetic fields   the series \eqref{eq:K0:s-0:B8} and \eqref{eq:K0:s1:B5} can be considered as the series of perturbation theory, but at high magnetic fields they begin to diverge and the description in terms of  magnetic polarizabilities becomes poorely defined. So we try to find the appropriate range of fields to obtain the dipole magnetic polarizability, as it was done for  vector $\rho$ mesons in our previous work \cite{Luschevskaya:2017}.
  \begin{figure}[htb]
\begin{center}
\includegraphics[width=6cm,angle=-90]{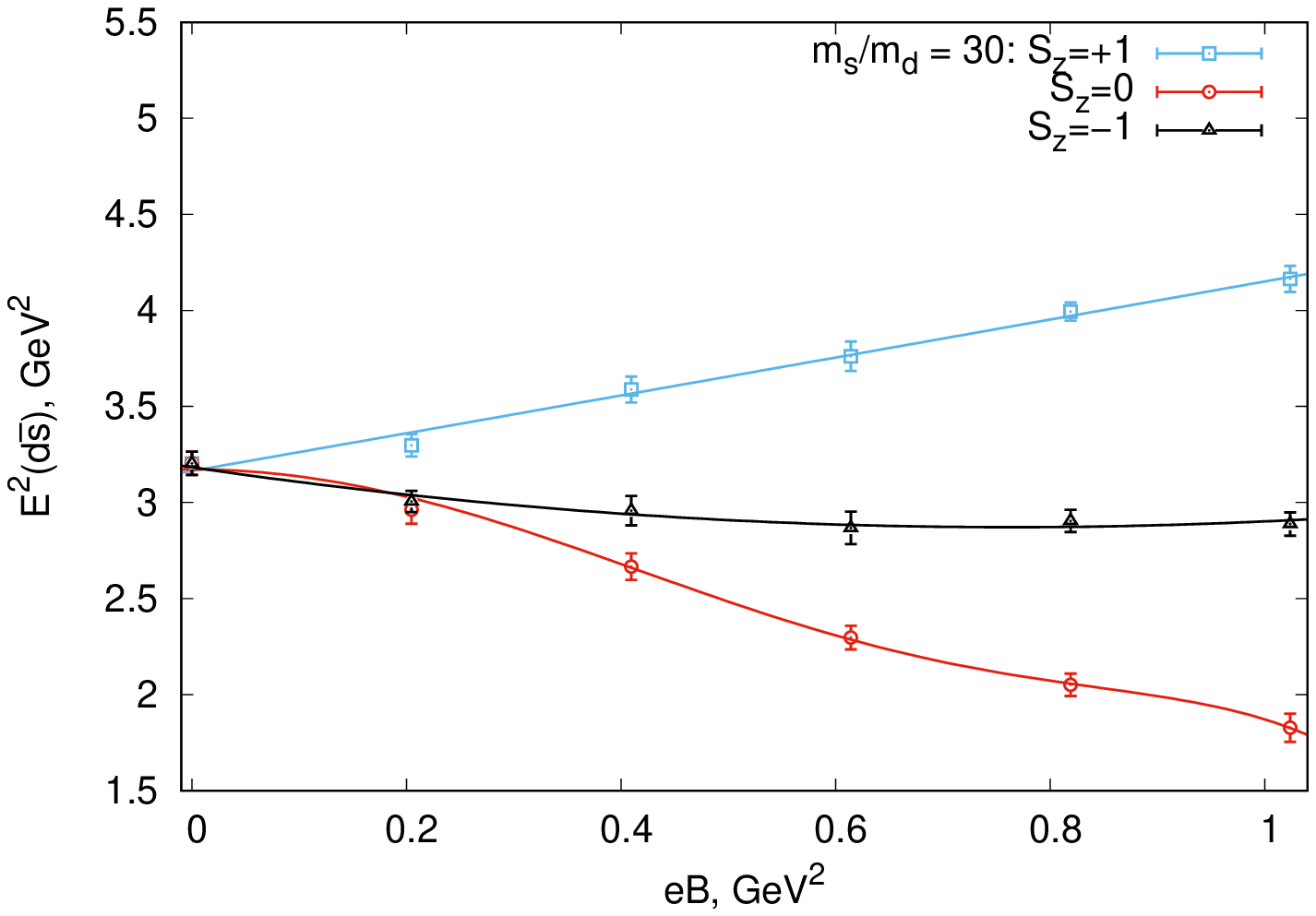}
\caption{The  energy squared of the  $K^{*0}$  meson for the various  meson spin projections on the magnetic field direction depending on the field value for the lattice volume $18^4$, lattice spacing $0.105\ \fm$, the pion mass $m_{\pi} = 367(8)\ \MeV$ and $m_s/m_d=30$.}
\label{Fig:K30_all_spins}
\end{center}
\end{figure}
\begin{figure}[htb]
\begin{center}
\includegraphics[width=6cm,angle=-90]{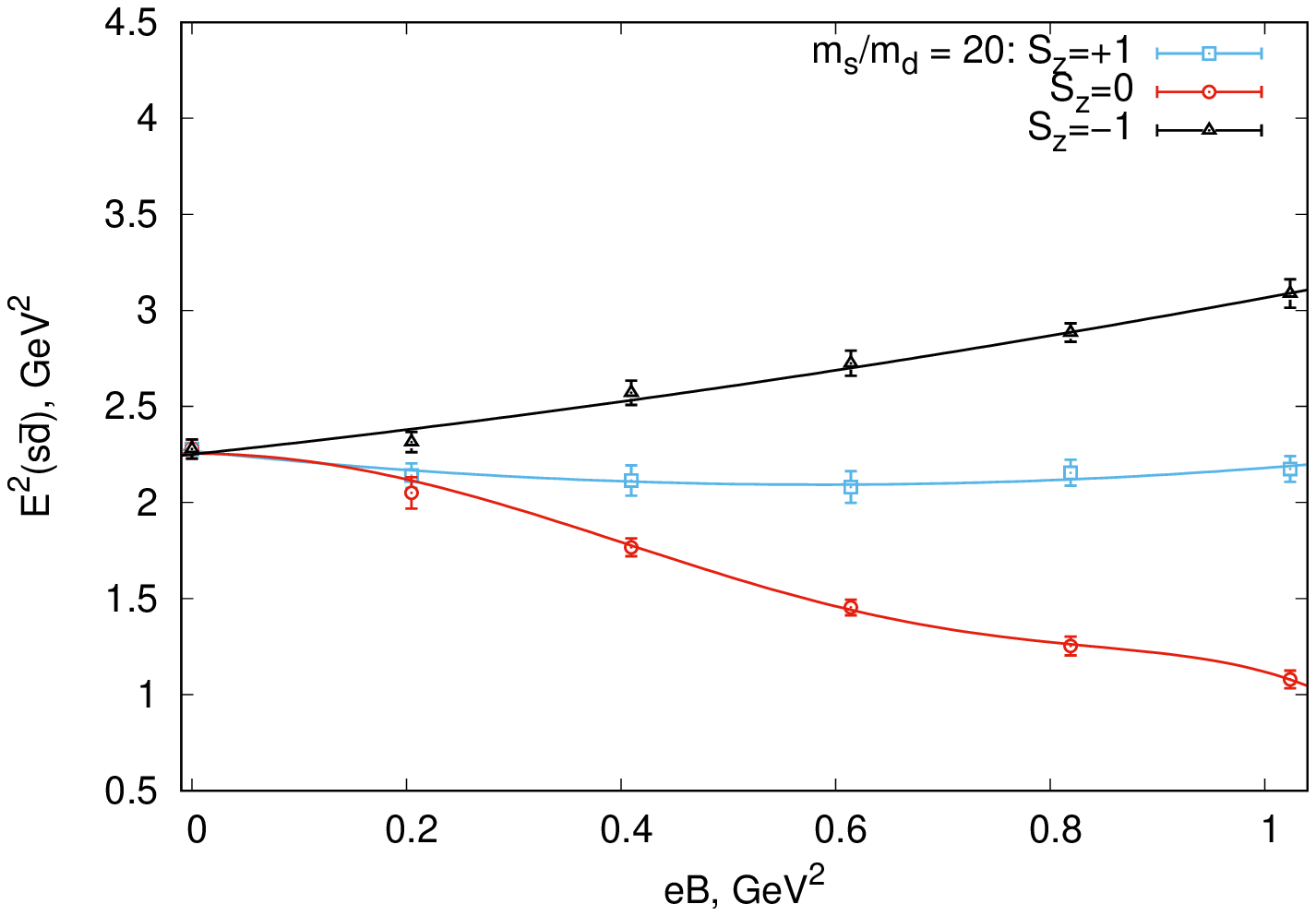}
\caption{The  energy squared of the $\bar{K}^{*0}$  meson for $S_z=-1,0$ and $+1$ spin projections on the magnetic field axis for the lattice volume $18^4$, lattice spacing $0.105\ \fm$,  pion mass $m_{\pi} = 367(8)\ \MeV$ and $m_s/m_d=20$.}
\label{Fig:K20_all_spins}
\end{center}
\end{figure}

In Fig.~\ref{Fig:K30_all_spins} we show the energy squared of the neutral vector $K^{*0}$ meson ($d\bar{s}$, $J=1$)   for  various meson spin projections  onto the direction of the external magnetic field for the ratio of  quark masses $m_s/m_d = 30$, the lattice volume  $18^4$, lattice spacing $0.105\ \fm$ and the pion mass equal $367(8)\ \MeV$. The points are the lattice data and the  lines correspond to the fits of these data.

We fit the energy squared for the  $S_z=0$  energy branch   by the  formula \eqref{eq:K0:s-0:B8}. It gives the best fit  at $eB \in [0:1.03]\,\GeV^2$, if we are limited  to the  terms of the sixth degree of  field  $\sim (eB)^6$. The mass   of the meson $m$, the magnetic dipole polarizabilty $\beta_m$, the hyperpolarizabilities $\beta_m^{h1}$ and $\beta_m^{h2}$ are the fit parameters. The energy of this spin component smoothly decreases with the increase of the field motivating further investigations at larger fields.

The lattice data for nonzero spin projections $S_z=\pm 1$ is fitted by the formula \eqref{eq:K0:s1:B5}.     At $eB \in [0:1.23]\,\GeV^2$ the best fit is obtained  taking into account the  terms up to the third order  $\sim (eB)^2$. The mass  $m$, the $g$-factor and the magnetic dipole polarizability $\beta_m$  are the fits parameters. The energy of the meson with the spin projection  $S_z=+1$  increases with the magnetic field, while the energy of the $S_z=-1$ spin component decreases very slightly at low fields and   tends to stay constant within the errors for these range of fields.

 At a given range of fields the choice of the fit, i.e. the number of terms in \eqref{eq:K0:s-0:B8} and \eqref{eq:K0:s1:B5},  is determined  by the best value of  $\chi^2$  and $p$-value of significance level.  Thus, we fitted   the available data at various field ranges including more terms of higher field degrees of the field  with an increase of the fitting range. At the same time,  we make sure that with an expansion of the fitting range  the results are in agreement with the ones obtained for shorter fitting intervals within the errors, see  Appendix \ref{app1}.

In Fig.~\ref{Fig:K20_all_spins} we show the energy squared of the vector meson $\bar{K}^{*0}$ ($s\bar{d}$)   for the smaller value of the ratio $m_s/m_d = 20$.  The energy branches with spins $S_z = -1$ and $S_z=+1$  exchanged  with respect to the case of $K^{*0}$ meson because of the positive value of the magnetic moment, which is opposite in sign to the magnetic moment of the $K^{*0}$ meson.   Now the energy of the meson with $S_z = -1$ increases, while the energy of the meson with $S_z = +1$  stays almost constant.

It can also be seen from a  comparison of
   Fig.~\ref{Fig:K30_all_spins} and    Fig.~\ref{Fig:K20_all_spins}, that for the smaller ratio of strange to light quark masses $m_s/m_d=20$ the difference between various energy branches is smaller,   the energy changes slower with the field value than for the $m_s/m_d = 30$ case. This behavior is consistent with the increase of    the  magnetic moment absolute value  with the strange quark mass, which is represented in the following Subsection.
Also   the  mass of the meson   in formula  \eqref{eq:K0:s-0:B8}  presents as a factor in the each term of the series. Therefore,  for the meson with a lower mass the decrease of its energy for zero spin component will be slower with the magnetic field if the magnetic polarizabilities do not strongly depend on the meson mass.

\subsection{Magnetic moment and magnetic dipole polarizability}
\label{subsec-42}

\ \ \ \ \ We  calculate  the ground state energy and the magnetic dipole polarizability $\beta_m$ of the vector $K^{*0}$ meson   for different ratios of the bare strange quark mass to the bare light quark mass $m_s/m_d$. For the $K^{*0}(\bar{K}^{*0})$ meson we have   $m_s/m_d = 20,30$. Then we decrease the strange quark mass ratio to study how the $\beta_m$ changes with the value of $m_s/m_d$.

   \begin{figure}[htb]
\begin{center}
\includegraphics[width=6cm,angle=-90]{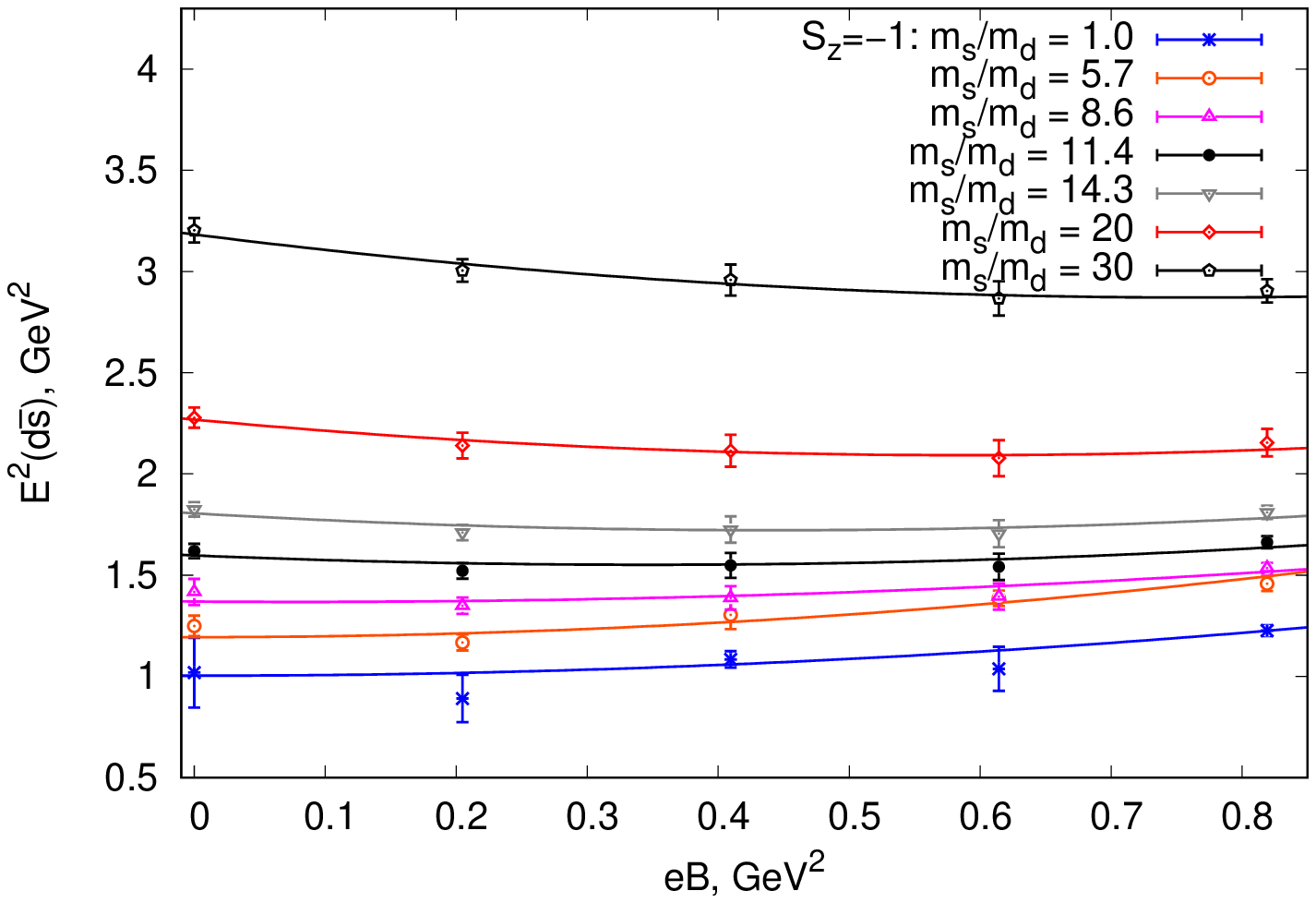}
\caption{The energy squared of   the $K^{*0}$  meson for the spin projection $S_z = -1$  depending on the magnetic field value  for the lattice volume $18^4$, the lattice spacing $0.105\ \fm$, the  pion mass $m_{\pi} = 367(8)\ \MeV$ and different values of $m_s/m_d$ ratio.}
\label{Fig:K0_S_-1}
\end{center}
\end{figure}

 \begin{table*}
\begin{ruledtabular}
\begin{tabular}{ccccccc}
  $m_s/m_u$  &  $g$-factor  & $\beta_m(\GeV^{-3})$  & $\chi^2$/n.d.f. & n.d.f. & p-value        & $eB(\GeV^2)$\\
\hline
    $1$      & -                    & $-0.026 \pm 0.006$   & $0.751$      & $3$    &   $0.522$    & $[0:0.82]$\\
   $5.7$     & -                    & $-0.033 \pm 0.003$  & $1.042$       & $4$    &    $0.384$   & $[0:1.03]$\\
    $8.6$     &  $-0.044\pm 0.106 $  & $-0.019\pm 0.005$     & $0.510$     & $4$  &    $0.728$      & $[0:1.23]$\\
   $11.4 $  &  $-0.265\pm 0.100$  & $-0.024 \pm 0.005$     & $0.628$      & $4$    & $0.642 $    & $[0:1.23]$\\
   $14.3 $   &  $-0.378 \pm 0.097$  & $-0.025 \pm 0.004$   & $0.535 $     &  $4$   &    $0.710$   & $[0:1.23]$\\
   $20$      &  $-0.599 \pm 0.076$  & $-0.027 \pm 0.003$   & $0.142$      & $4$    &   $0.966$    & $[0:1.23]$\\
   $30$      &  $-0.816 \pm 0.103$  & $-0.024 \pm 0.003$   & $0.244 $     & $4$    &   $0.913$    & $[0:1.23]$\\
\end{tabular}
\caption{The   $g$-factor value and the magnetic dipole polarizability $\beta_m$ of the $d\bar{s}$ state  for $S_z=-1$  with other fit parameters  obtained for different mass ratios $m_s/m_d$. The data were fitted at $m_s/m_d \leqslant 5.7$  and $m_s/m_d \geqslant 8.6$ by the formulas \eqref{eq:K0:s1:B2} and \eqref{eq:pt3:K0:s1:B2} respectively.  The ranges of the magnetic fields used for the fit are shown in the last column.}
\label{Table:beta:K0_S-1}
\end{ruledtabular}
\end{table*}

In Fig.~\ref{Fig:K0_S_-1} we depict the energy squared of the meson for the $S_z=-1$  spin projection  for the  lattice spacing $0.105\ \fm$, lattice volume $18^4$, pion mass $m_{\pi} = 367(8)\ \MeV$  and different values of $m_s/m_d$ ratio. The points correspond to the lattice data. The lines represent the fits of the lattice data by theoretical curves.

We observe that the the magnetic moment of the vector $d\bar{s}$ state is negative in sign and diminishes with    the growth of the strange bare quark mass. It has to be equal to zero at $m_s/m_d=1$. From  Table \ref{Table:beta:K0_S-1_comparison}  in Appendix \ref{app1} one can see that even at $m_s/m_d=5.7$ the $g$-factor  is zero within the error bars, so at $m_s/m_d \leqslant 5.7$ we fit the data by the formula
  \begin{equation}
E^2 = m^2   - 4 \pi m \beta_m (eB)^2,
\label{eq:K0:s1:B2}
 \end{equation}
  where the mass $m$ of the meson at zero field  and the  magnetic dipole polarizabilty $\beta_m$ are the fit parameters, which are collected   in Table \ref{Table:beta:K0_S-1}.   We do not consider  the term  $\sim(eB)^3$  with hyperpolarizability  because its contribution is insignificant within the errors for the field ranges considered.

At higher masses of the heavier quark, i.e. at $m_s/m_d  \geqslant 8.6$  we cannot neglect the contribution of the magnetic moment and include the corresponding linear term into the fit. At lower fields $eB\lesssim 0.82\,\GeV^2$ the better fits are obtained with the linear law
  \begin{equation}
E^2 = m^2 -g(eB).
\label{eq:K0:s1:B1}
 \end{equation}

If  we increases the range of fields used for the fit, consider the values $eB\geqslant0.9\,\GeV^2$, the magnetic moment is overestimated    and the quality of the fit getting worse, see again  Table \ref{Table:beta:K0_S-1_comparison} in Appendix \ref{app1}. However, if we add to the fit the  term  with dipole polarizability $\sim (eB)^2$ and enlarge the field range for the fit, then  we get the  $g$-factor value comparable with that obtained at low fields using 2-parametric fit \eqref{eq:K0:s1:B1}.  This suggests that the terms with higher field degrees begin to make a larger contribution to the meson energy at higher fields.
 The fits have been chosen so that they gives the best p-value of significance level and $\chi^2/n.d.f.$ value at the corresponding field range.

So, for the masses $m_s/m_d  \geqslant 8.6$  the better fits are obtained at $eB\in[0:1.23]\,\GeV^2$ using 3-parametric fits
  \begin{equation}
E^2 = m^2 - g(eB) - 4 \pi m \beta_m (eB)^2,
\label{eq:pt3:K0:s1:B2}
 \end{equation}
 where the mass $m$ of the meson at zero field, the $g$-factor value and the  magnetic dipole polarizabilty $\beta_m$ are shown in  Table \ref{Table:beta:K0_S-1}. At such field range the  3-parametric fit  gives the $g$-factor values which are consistent with ones obtained at $eB\in[0;0.41]\,\GeV^2$ from 2-parametric fit \eqref{eq:K0:s1:B1}.
 
 Our $g$-factor value of the $K^{*0}$ meson agrees in sign with the result of the previous work $g(K^{*0})=-0.183$ \cite{Simonov:2013b}. As for the quantitative discrepancies, this requires further extrapolations to a   physical mass of the pion and investigations of the lattice volume and lattice spacing effects.

The magnetic dipole polarizability is negative in sign and seemingly does not depend on the  strange quark  mass within the errors.
The value of the  magnetic dipole polarizability $\beta_m = -0.026\pm 0.006\ \GeV^{-3}$ obtained at  $m_s/m_d =1$ corresponds to the $d\bar{d}$  state and agrees in sign and   order of magnitude with the value of  the  magnetic polarizability obtained for the  neutral vector $\rho^0$ meson  \cite{Teryaev:Tensorpolar}. The absolute value of the magnetic dipole polarizability of $\rho^0$ meson is larger  than for $d\bar{d}$ state, because the $\rho^0$ meson also contains $u$-quark making it  more susceptible to the influence of the external strong magnetic field. Previously, for the pion mass   $m_{\pi}=395 \pm 6 \ \MeV$  we get the value $\beta_m = -0.11\pm 0.03\ \GeV^{-3}$,  the lattice calculations at $m_{\pi}=541 \pm 3 \ \MeV$ give the value    $\beta_m =-0.07\pm 0.02\ \GeV^{-3}$ \cite{Teryaev:Tensorpolar}.

 \begin{figure}[htb]
\begin{center}
\includegraphics[width=6cm,angle=-90]{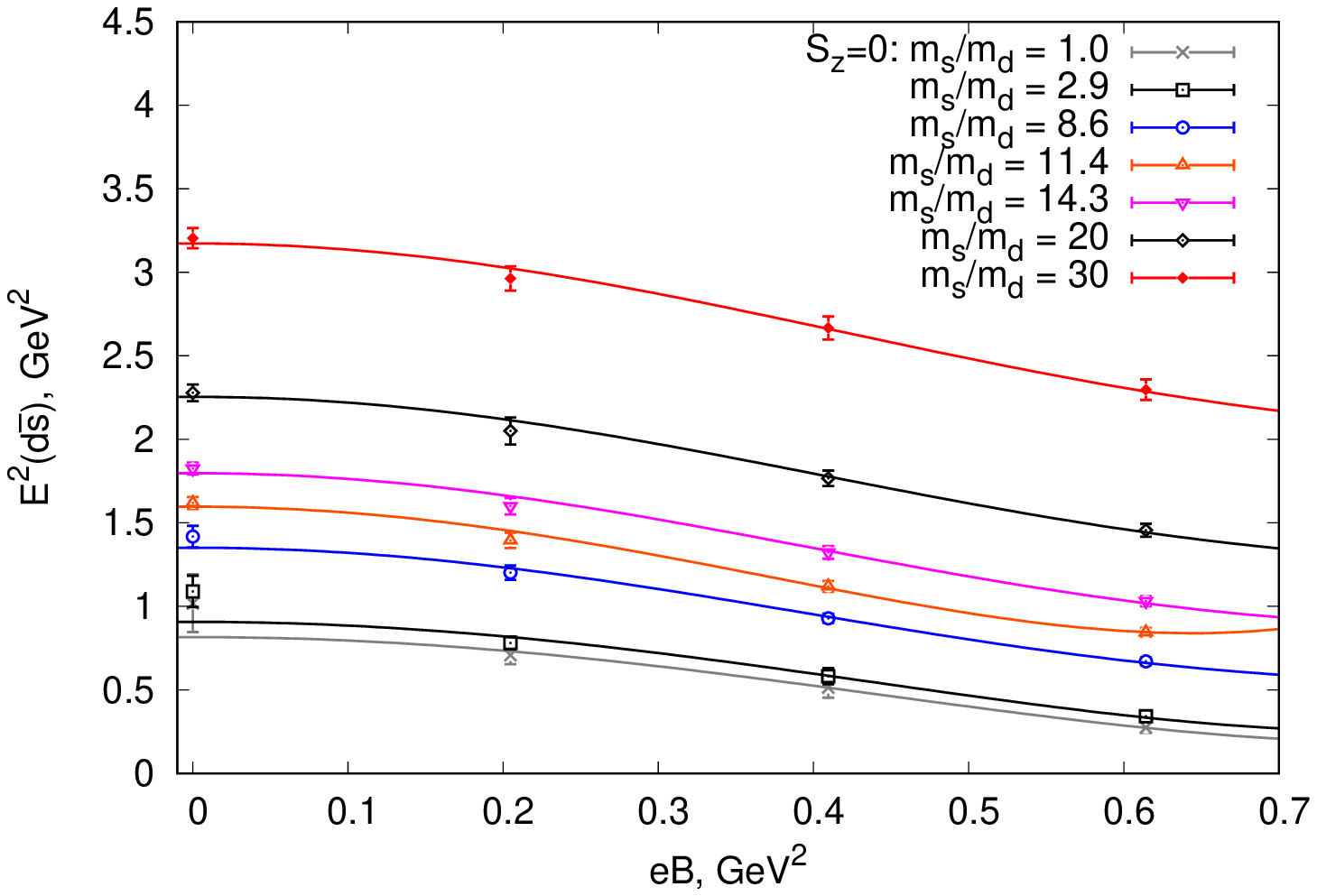}
\caption{The energy squared of   the $K^{*0}$  meson for the spin projection $S_z = 0$  depending on the magnetic field    for the lattice volume $18^4$, the lattice spacing $0.105\ \fm$, the  pion mass $m_{\pi} = 367(8)\ \MeV$ and different values of $m_s/m_d$ ratio.}
\label{Fig:K0_S_0}
\end{center}
\end{figure}

 \begin{table*}
\begin{ruledtabular}
\begin{tabular}{cccccccc}
  $m_s/m_d$  &  $\beta_m(\GeV^{-3})$  & $\beta_m^{h1}(\GeV^{-5})$ & $\beta_m^{h2}(\GeV^{-7})$ & $\chi^2$/n.d.f. & n.d.f.  & p-value &  $eB(\GeV^2)$\\
\hline
   $1$     & $0.185 \pm 0.022$    & $-0.155 \pm 0.030$   & -                  & $0.784$  &   $2$  &   $0.456$    & $[0:0.82]$\\
   $2.9$   & $0.188 \pm 0.032$    & $-0.162 \pm 0.044$   & -                  & $2.346$  &   $2$  &    $0.096$   & $[0:0.82]$\\
   $5.7$   & $0.232 \pm 0.044 $   &  $-0.328 \pm 0.100$   & $0.158 \pm 0.058$ & $3.759$  &  $2$   &    $0.023$   &  $[0:1.03]$\\
   $8.6$    &  $0.212 \pm 0.019$  & $-0.278 \pm 0.041$    & $0.127 \pm 0.024$ & $0.861 $    &  $2$   & $0.423 $ & $[0:1.03]$\\
   $11.4$   & $0.230 \pm 0.039  $  & $-0.277 \pm 0.098$  & -       &    $1.998$           & $1$    & $0.157$ & $[0:0.62]$\\
   $14.3$   &  $0.206 \pm 0.019$  & $-0.167 \pm 0.047$      & $0.122 \pm 0.029$  &    $2.198$  & $2$ & $0.311$ & $[0:1.03]$\\
   $20$    &  $0.187\pm 0.017$ & $-0.233\pm0.044$     & $ 0.107 \pm 0.028 $ & $0.530$    & $2$    & $0.588$    & $[0:1.03]$\\
  $30$     & $0.167\pm0.019$   & $-0.198 \pm 0.049$   &  $0.089 \pm 0.031$  & $0.501$    & $2$    &  $0.606 $  & $[0:1.03]$\\
\end{tabular}
\caption{The magnetic dipole polarizability $\beta_m$, the hyperpolarizability of the first order $\beta_m^{h1}$ and the magnetic hyperpolarizability of the second order $\beta_m^{h2}$ of the vector $d\bar{s}$ state for $S_z=0$  with other fit parameters. At $m_s/m_d=1,2.9,11.4$ the better fits were obtained using the formula \eqref{eq:K0:s-0:B4}. For other $m_s/m_d$ values
  the formula  \eqref{eq:K0:s-0:B6} gives the better fits of the lattice data and allows to roughly estimate the magnetic hyperpolarizability of the second order $\beta_m^{h2}$. The range of the magnetic fields used for the fit is shown in the last column.}
\label{Table:beta:K0_S0}
\end{ruledtabular}
\end{table*}

In Fig.~\ref{Fig:K0_S_0} we show the energy squared of $K^{*0}$ meson corresponding to the  $m_s/m_d=20,30$  for the case  of antiparallel quark spins, i.e. $S_z=0$, hypothetical states with $m_s/m_d<20$ are also shown.  We observe the   decrease of energy  with  the magnetic field value  for  all  the values of  $m_s/m_d$. The terms of higher powers of the magnetic field begin to contribute significantly at relatively low fields, so for   $eB\in[0:0.9]\,\GeV^2$ we use  3-parametric  fit
 \begin{equation}
E^2 = m^2 -4 \pi m \beta_m (eB)^2 - 4 \pi m \beta_m^{h1} (eB)^4,
\label{eq:K0:s-0:B4}
\end{equation}
where the meson mass $m$, the magnetic dipole polarizability $\beta_m$ and hyperpolarizability of the first order $\beta_m^{h1}$ are the fit parameters.

For  $eB\in[0:1.1\,\GeV^2]$  4-parametric fit
  \begin{eqnarray}
E^2 =&&m^2 -4 \pi m \beta_m (eB)^2 \nonumber \\
 &&- 4 \pi m \beta_m^{h1} (eB)^4 -  4 \pi m \beta_m^{h2}(eB)^6
\label{eq:K0:s-0:B6}
\end{eqnarray}
gives the best p-value of significance level.
  When we enlarge the field range considered, the terms with the hyperpolarizabilities $\beta_m^{h1}$ and $\beta_m^{h2}$ begin to contribute to the meson energy. This  underestimates the  magnitude of $\beta_m$ as one can see from Table \ref{Table:beta:K0_S0_comparison}  so we   include the  terms $\sim (eB)^4$ and $\sim (eB)^6$ into a consideration and choose the range of fields to get the most significant fit for the each value of $m_s/m_d$ ratio.
The obtained values of magnetic polarizabilities    are represented   in Table \ref{Table:beta:K0_S0}.
One can see that the magnetic dipole polarizability
does not depend on the bare strange quark mass within the errors.

\begin{figure}[htb]
\begin{center}
\includegraphics[width=6cm,angle=-90]{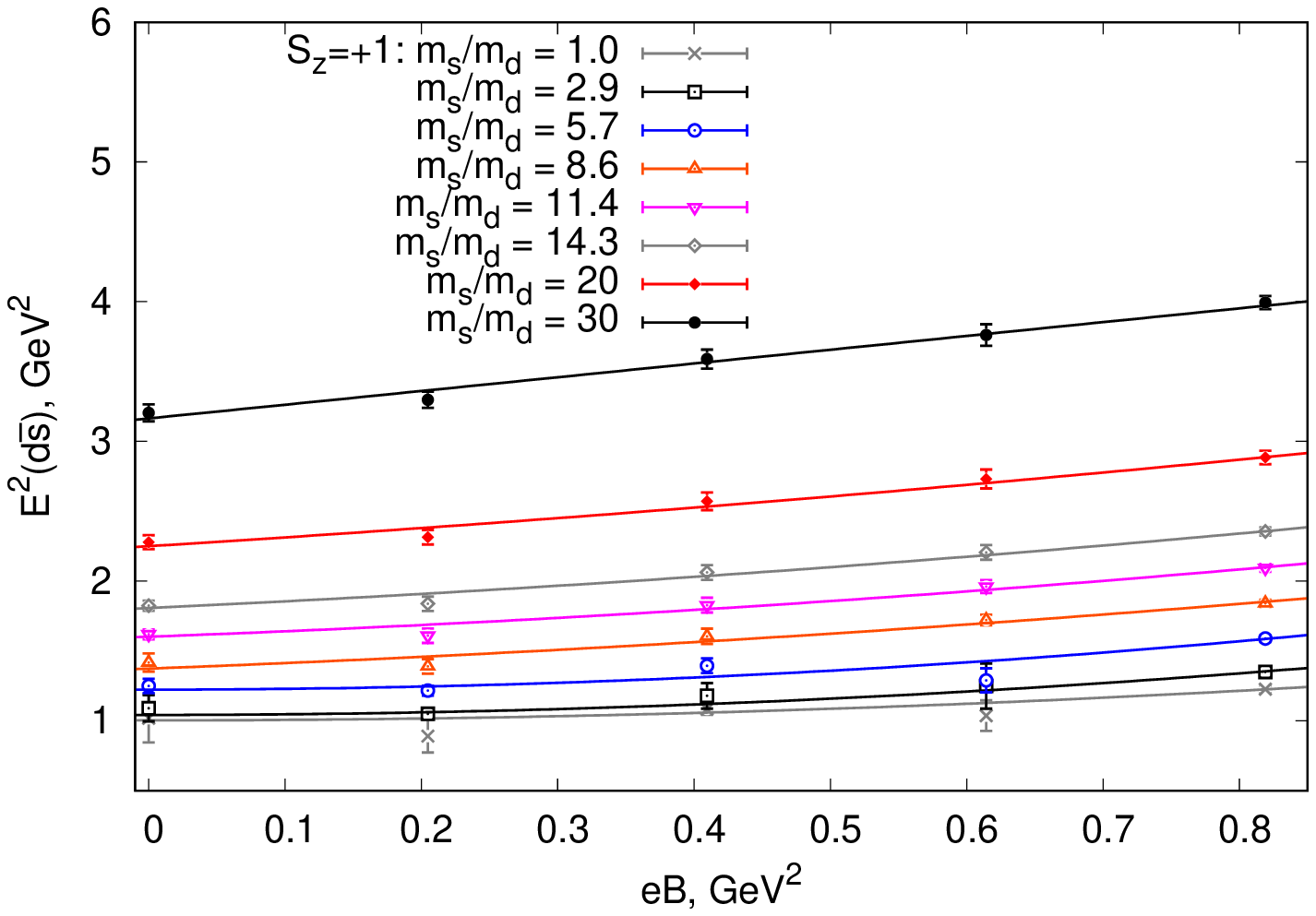}
\caption{The  energy squared of the $K^{*0}$  meson for the spin projection $S_z = +1$  for different values of  the  $m_s/m_d$ ratio depending on the magnetic field with the corresponding fits of these data by the formula \eqref{eq:K0:s1:B2}. The calculations were performed for the lattice volume $18^4$, lattice spacing $0.105\ \fm$ and the pion mass $m_{\pi} = 367(8)\ \MeV$.}
\label{Fig:K0_S_+1}
\end{center}
\end{figure}

\begin{table*}
\begin{ruledtabular}
\begin{tabular}{ccccccc}
 $m_s/m_d$ & $g$-factor & $\beta_m (\GeV^{-3})$    & $\chi^2$/n.d.f. & n.d.f. & p-value & $eB(\GeV^2)$\\
\hline
   $1$    &   -                  & $-0.026 \pm 0.006$        & $0.751$      & $3$    &   $0.522$    & $[0:0.82]$\\
   $2.9$  &  -                   & $-0.036 \pm 0.002$        & $0.216 $     & $4$    &   $0.930$    & $[0:1.03]$\\
   $5.7$   &  -                   & $-0.039 \pm 0.006$        & $1.952$      & $3$    &   $0.119$    & $[0:0.82]$\\
   $8.6$   &  $-0.367 \pm 0.198$  & $-0.018 \pm 0.012$        & $0.916$      & $4$    &   $0.453$    & $[0:1.23]$\\
   $11.4$  &  $-0.357 \pm 0.153$  & $-0.019 \pm 0.010$        & $0.866$      & $4$    &   $0.484$    & $[0:1.23]$\\
   $14.3$  &  $-0.449 \pm 0.143$  & $-0.016 \pm 0.009$        & $0.741$      & $4$    &   $0.564$    & $[0:1.23]$\\
   $20$    &  $-0.606 \pm 0.173$  & $-0.011 \pm 0.009$        & $0.650$      & $4$    &   $0.627$    & $[0:1.23]$\\
  $30$     &  $-0.981 \pm 0.172$  & $-0.0002\pm 0.007$        & $0.565$      & $4$    &   $0.688$     & $[0:1.23]$\\
\end{tabular}
\caption{The  $g$-factor value and the magnetic dipole polarizability $\beta_m$  of the vector $K^{*0}$  meson for the case of $S_z=+1$ spin projection with other fit parameters  for different values of  $m_s/m_d$ ratio. The  range of fields  used for the fit is shown in the last column.}
\label{Table:beta:K0_S+1}
\end{ruledtabular}
\end{table*}

In Fig.~\ref{Fig:K0_S_+1} we show the energy squared of the  meson with $S_z=+1$. The lines correspond to the fits of  lattice data by    2-parametric fit \eqref{eq:K0:s1:B2} for $m_s/m_d\leq 5.7$  and  3-parametric fit \eqref{eq:pt3:K0:s1:B2} for $m_s/m_d\geqslant8.6$.  The choice of the fits for the considered range of  fields is determined by the best $\chi^2/n.d.f.$ and $p$-value.  In   Table \ref{Table:beta:K0_S+1_comparison} of Appendix \ref{app1} it is shown how  the $g$-factor  and dipole polarizability $\beta_m$ depend on the range of the magnetic fields  chosen for the fit. The best results were included in Table \ref{Table:beta:K0_S+1} and shown in Fig.~\ref{Fig:K0_S_+1}. From \ref{Table:beta:K0_S+1} one can conclude that the
  $g$-factor of the vector $d\bar{s}$ state decreases with the increase of the strange quark mass as it was found from the $S_z=+1$ energy branch. The values of the magnetic moment of the $K^{*0}$ meson obtained from $S_z=-1$ and $S_z=+1$ energy branches agree with each other within the errors.

The magnetic dipole polarizability for $S_z=+1$ energy component  increases from some negative value $-0.026\pm 0.006\ \GeV^{-3}$ at $m_s/m_d=1$ to zero value at $m_s/m_d=30$. Due to parity conservation, we expect equal values  of dipole polarizability for the $S_z=-1$ and $S_z=+1$ energy component.  This behavior may be a result of the lattice spacing or lattice volume effects and needs further investigations.

  \subsection{Tensor polarizability of $K^{*0}$ and $\bar{K}^{*0}$  mesons}
  \label{subsec-43}
  \begin{table*}
\begin{ruledtabular}
\begin{tabular}{ccccc}
  $m_s/m_u$  &  $\beta_{S=+1} (\GeV^{-3})$  & $\beta_{S=-1} (\GeV^{-3})$  & $\beta_{S=0} (\GeV^{-3})$ & $\beta_t$ \\
\hline
  $1$       &  $-0.026 \pm 0.006 $   &  $-0.026 \pm 0.006$   &  $0.185 \pm 0.022$   &  $ -3.17 \pm 0.33$   \\
  $5.7$     &  $-0.039 \pm 0.006$    &  $-0.033 \pm 0.003$   &  $0.232 \pm 0.044$   &  $ -3.35 \pm 0.41$   \\
  $8.6$     &  $-0.018\pm 0.012$    &  $-0.019 \pm 0.005$   &  $0.212 \pm 0.019$   &  $ -3.63 \pm 0.28$   \\
  $11.4$    &  $-0.019 \pm 0.010$    &  $-0.024 \pm 0.005$   &  $0.230 \pm 0.039$   &  $ -3.69 \pm 0.26$   \\
  $14.3$    &  $-0.016 \pm 0.009$    &  $-0.025  \pm 0.004$   &  $0.206 \pm 0.019$   &  $ -2.75 \pm 0.24$   \\
  $20$      &  $-0.011 \pm 0.009$    &  $-0.027  \pm 0.003$   &  $0.187 \pm 0.017$   &  $ -2.77 \pm 0.26$   \\
  $30$      &  $-0.0002\pm 0.007$    &  $-0.024  \pm 0.003$   &  $0.167 \pm 0.019$   &  $ -2.51 \pm 0.20$   \\
\end{tabular}
\caption{The tensor polarizability of the  $K^{*0}$ and $\bar{K}^{*0}$  shown in the fifth column obtained for various values of $m_s/m_d $ ratio for the pion mass   $m_{\pi}=367\pm 8 $, lattice volume $18^4$ and the lattice spacing $a=0.105\ \fm$.}
\label{Table:tensorpolar:K0}
\end{ruledtabular}
\end{table*}

\ \ \ \ The energy of the $K^{*0}$ meson with $S_z=+1$ increases, while the energy for zero spin component $S_z=0$   diminishes versus the  field value.
At low magnetic field and $m_s/m_d\geqslant 20$ the energy  behavior of the $S_z=-1$ energy component is determined by the magnetic moment value, so the energy decreases weakly for $eB\lesssim 0.3\ \GeV^2$ and at larger fields stays almost constant or slowly grows   within the errors. For small  values of  $m_s/m_d\lesssim8.6$ ratio the energy grows with the magnetic field slowly, for other $m_s/m_d$ values the energy is almost constant within the errors at the range of fields considered.

The low energy is less profitable than the high one, therefore   meson states with $S_z=0$ corresponding to the longitudinal polarization will prevail in non-central heavy ion collisions.
The   vector dominance principle states that vector mesons can  convert  through intermediate resonances  to the virtual photons which decay electromagnetically to dileptons.  Therefore, the spin structure of resonances   defines the  anysotropy of  emitted dileptons.

The   differential cross section of dileptons has the following form
\begin{equation}
 \frac{d\sigma}{dM^2 d\cos\theta}=A(1+B\cos^2 \theta),
 \label{cross_sec}
\end{equation}
where $\theta$ is the angle between the momenta of a lepton and a virtual photon, $M$ is the energy of the lepton pair in their rest frame.   The asymmetry coefficient is defined by the relation
 \begin{equation}
 B =\frac{\gamma_{\perp}-\gamma_{\parallel}}{\gamma_{\perp}+\gamma_{\parallel}},
\label{asymm}
\end{equation}
where the  $\gamma_{\perp}$ and $\gamma_{\parallel}$  are the contributions of the transverse and
longitudinal polarizations  of a virtual intermediate photon.
The   vector $K^{*0}$ and $\bar{K}^{*0}$ are   the sources of  the virtual photons. We perform the estimation of the asymmetry factor for such processes.

The polarization tensor of the vector particle in Cartesian basis reads
\begin{equation}
P_{ij}=\frac{3}{2}\langle s_i s_j+s_j s_i \rangle -2 \delta_{ij}.
\end{equation}
In  the    probability language the $P_{33}$  component can be written  by the following way
\begin{eqnarray}
P_{33}= &&3(w_{S_z=+1}+w_{S_z=-1}) \nonumber \\
 &&-2(w_{S_z=+1}+w_{S_z=-1}+w_{S_z=0}),
\label{asymmetry_coef}
\end{eqnarray}
where  $w_{S_z=+1}$, $w_{S_z=-1} $ and $w_{S_z=0}$ are the probabilities that the vector meson has a spin projection equal to $+1$, $-1$ and $0$ respectively. Since   $w_{S_z=+1}+w_{S_z=-1}+w_{S_z=0}=1$, it follows from \eqref{asymmetry_coef}  that  $P_{33}=1-3w_{S_z=0}$ and $-2 \leq P_{33}\leq 1$. The tensor polarizability reveals how the the magnetic field affect the   spin states of $K$ meson and its decay to lepton pair, see \cite{Terayev}.

For the transverse polarization of the $K^{*0}$ and $\bar{K}^{*0}$ mesons   $B=1$  and for the longitudinal  polarization $B=-1$.
Previously we introduce the  tensor polarizability in the following form
 \begin{equation}
 \beta_{t}=\frac{\beta_{S_z=+1}+\beta_{S_z=-1}-2\beta_{S_z=0}}{\beta_{S_z=+1}+\beta_{S_z=-1}+\beta_{S_z=0}}.
 \end{equation}
 supposing that the $B$ coefficient depends on this quantity in a hot media. In a strong magnetic field the spin meson tends to align along the field axis, but the temperature leads to the spin flips in a hot media.

 The values of tensor polarizability $\beta_{t}$ for the vector $K^{*0}$ and $\bar{K}^{*0}$  mesons  are presented in Table \ref{Table:tensorpolar:K0}. The spin components $S_z=+1$ and $S_z=-1$ of the $\bar{K}^{*0}$  meson changes places  with respect to the  $K^{*0}$ meson so the values $\beta_{t}$ are the same for the vector $K^{*0}$ and $\bar{K}^{*0}$  mesons.
 The large negative values of $\beta_t$ indicate that the longitudinal polarization dominates for the decays of these mesons.
 The dileptons are mainly emitted in the directions  close  to  the  perpendicular ones  to  the  magnetic  field  axis.
 The absolute value  of $\beta_t$ diminishes with the increase of $m_s/m_d$ ratio, however, the errors are sufficiently large.
 Probably, this may an indication that this effect  is weaker for  the heavier vector mesons.

\section{Vector  $K^{*\pm}$ mesons}
\label{sec-5}

\subsection{Energy}
\label{subsec-51}

\begin{figure}[htb]
\begin{center}
\includegraphics[width=6cm,angle=-90]{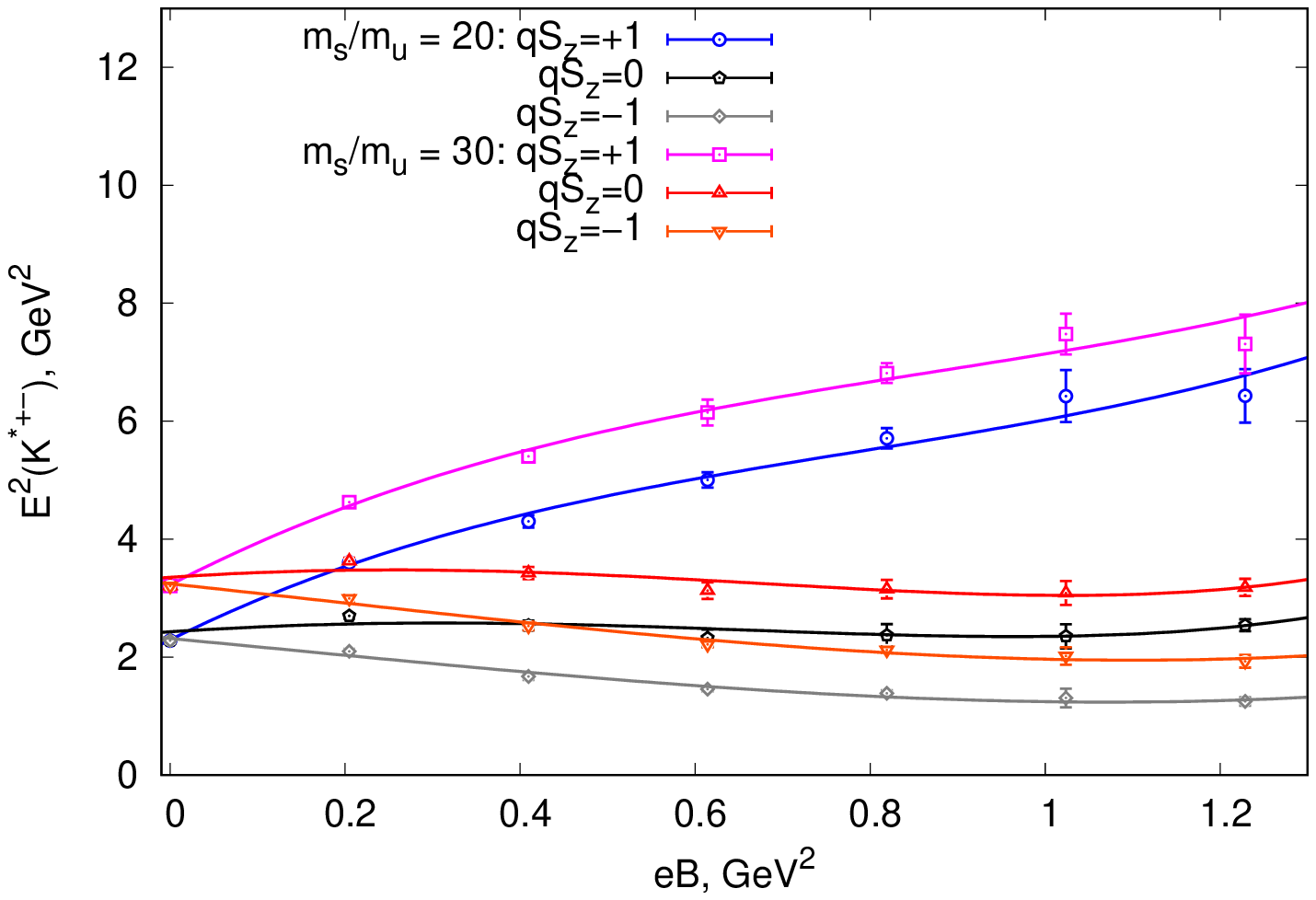}
\caption{The energy squared of the  vector $K^{*\pm}$ meson for various spin projections on  the external magnetic field. The calculations were performed at the lattice volume $18^4$, lattice spacing $a=0.105\ \fm$ and the pion mass $m_{\pi} = 367\pm 8 \ \MeV$.}
\label{Fig:K20K30_q1}
\end{center}
\end{figure}

The energies squared of the vector $K^{*\pm}$ mesons are represented in Fig.~\ref{Fig:K20K30_q1} for the values   $m_s/m_u=20$ and $30$, where $m_s$ is the bare strange quark mass and $m_u$ is the bare up-quark mass.    The energy of the charged vector $K$ meson with  $qS_z=+1$ increases, while the energy for the  $qS_z=-1$ case decreases at low field values and shows a constant behavior  at the magnetic fields from $1 \GeV^2$.

\begin{figure}[htb]
\begin{center}
\includegraphics[width=6cm,angle=-90]{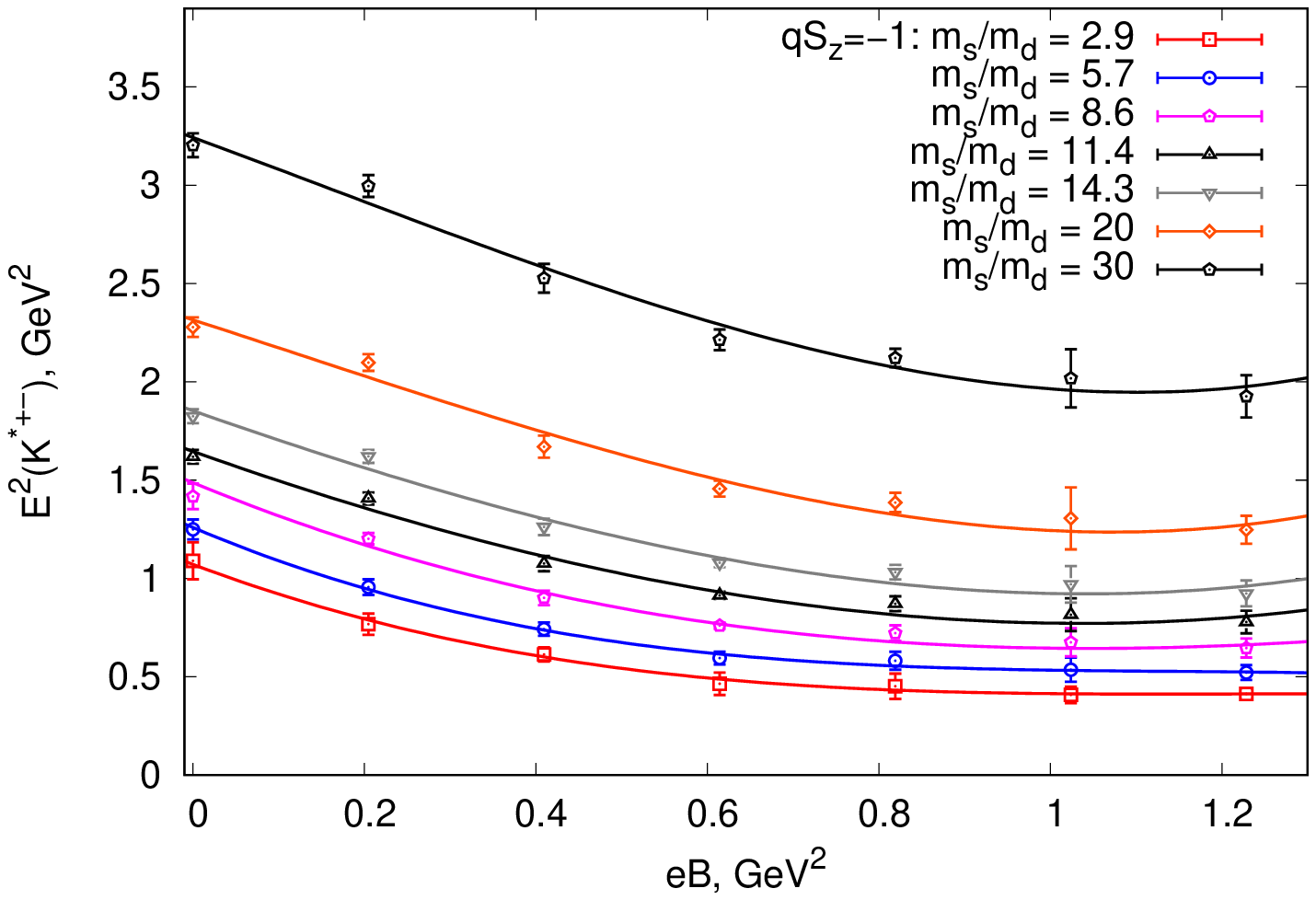}
\caption{The energy squared of the  vector $K^{*\pm}$ meson for $qS_z=-1$ spin projections on the external magnetic field for the same lattice parameters as   in Fig.~\ref{Fig:K20K30_q1}.}
\label{Fig:K_g_beta_plato}
\end{center}
\end{figure}

 In  Fig.~\ref{Fig:K_g_beta_plato} we depict the energy branch with $qS_z=-1$ for different values of the $m_s/m_d$ ratio. We see that with the growth of the strange quark mass the energy at the same magnetic field  increases that means the absense of the tachyon modes in QCD for these heavy-light quark states.

\subsection{Magnetic moments of the vector $K^{*\pm}$ mesons}
\label{subsec-52}

\begin{figure}[htb]
\begin{center}
\includegraphics[width=6cm,angle=-90]{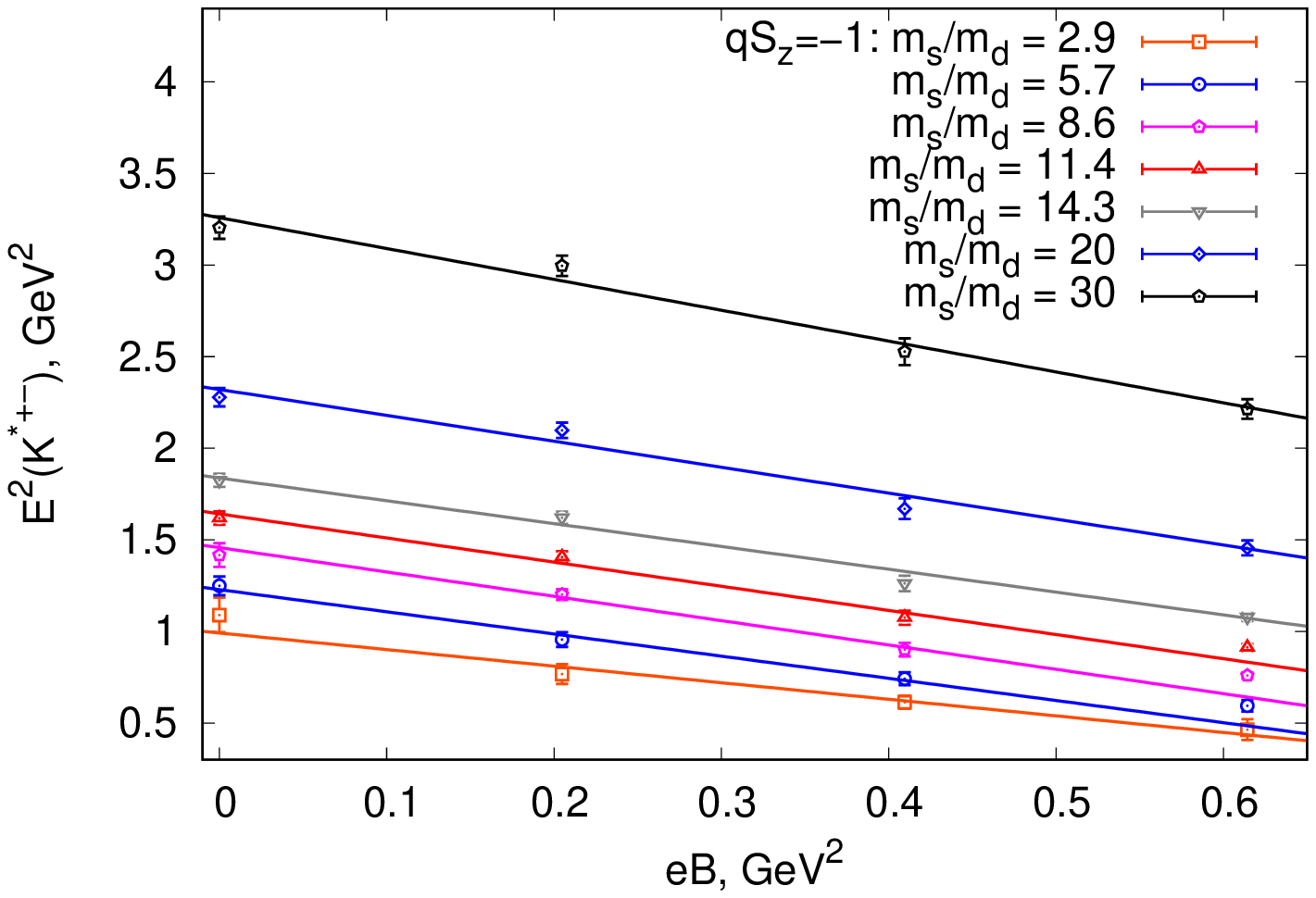}
\caption{The energy squared of the  vector $K^{*\pm}$ meson for $qS_z=-1$ spin projections on the external magnetic field for different ratios of the bare strange quark mass to the light quark mass.}
\label{Fig:K_g_factor}
\end{center}
\end{figure}

We estimate the Lande the g-factor of the vector $K^{*\pm}$  meson from the lattice data and explore its dependence from the  $m_s/m_d$ value. The magnitude of g-factor characterizes the magnetic moment of the particle in its natural magnetons. The  $K^{*\pm}$  mesons consists of strongly interacting   $u$, $s$ quarks and gauge fields.   It is hard to measure the magnetic moments of short lived particles like the $K^{*\pm}$ mesons, but their determination is of interest because it allows to explore the contribution of QCD effects.

  As one can see from the formula \eqref{eqLL} for Landau levels, the energy of a charged pointlike particle has a linear dependence from the magnetic field. At   large magnetic fields this formula gets nonlinear corrections. At the relatively low fields, i.e. at $eB\in[0:0.5]\ \GeV^2$ for $m_s/m_u\leq 11.4$, at $eB\in[0:0.8]\ \GeV^2$  for $m_s/m_u \geq 14.3$,   one can neglect these nonlinear effects and fit the lattice data by the formula \eqref{eqLL}.

To find the   $g$ - factor we use the low energy branch for which $q S_z =-1$, because it reveals much lower errors than the upper one with $qS_z = +1$, as it was done in our previous work \cite{Luschevskaya:2017}. Therefore we fit the lattice data for $K^{*+}$ with $S_z=-1$  or $K^{*-}$ with $S_z=+1$ by the formula
  \begin{equation}
E^2 =  |eB| + g(eB) + m^2,
\label{eq:K:g-factor}
 \end{equation}
 where $m$ and  $g$ are the fit parameters.

 In Fig.\ref{Fig:K_g_factor} the lattice data are shown by points   together with their linear fits.
 \begin{figure}[htb]
\begin{center}
\includegraphics[width=6cm,angle=-90]{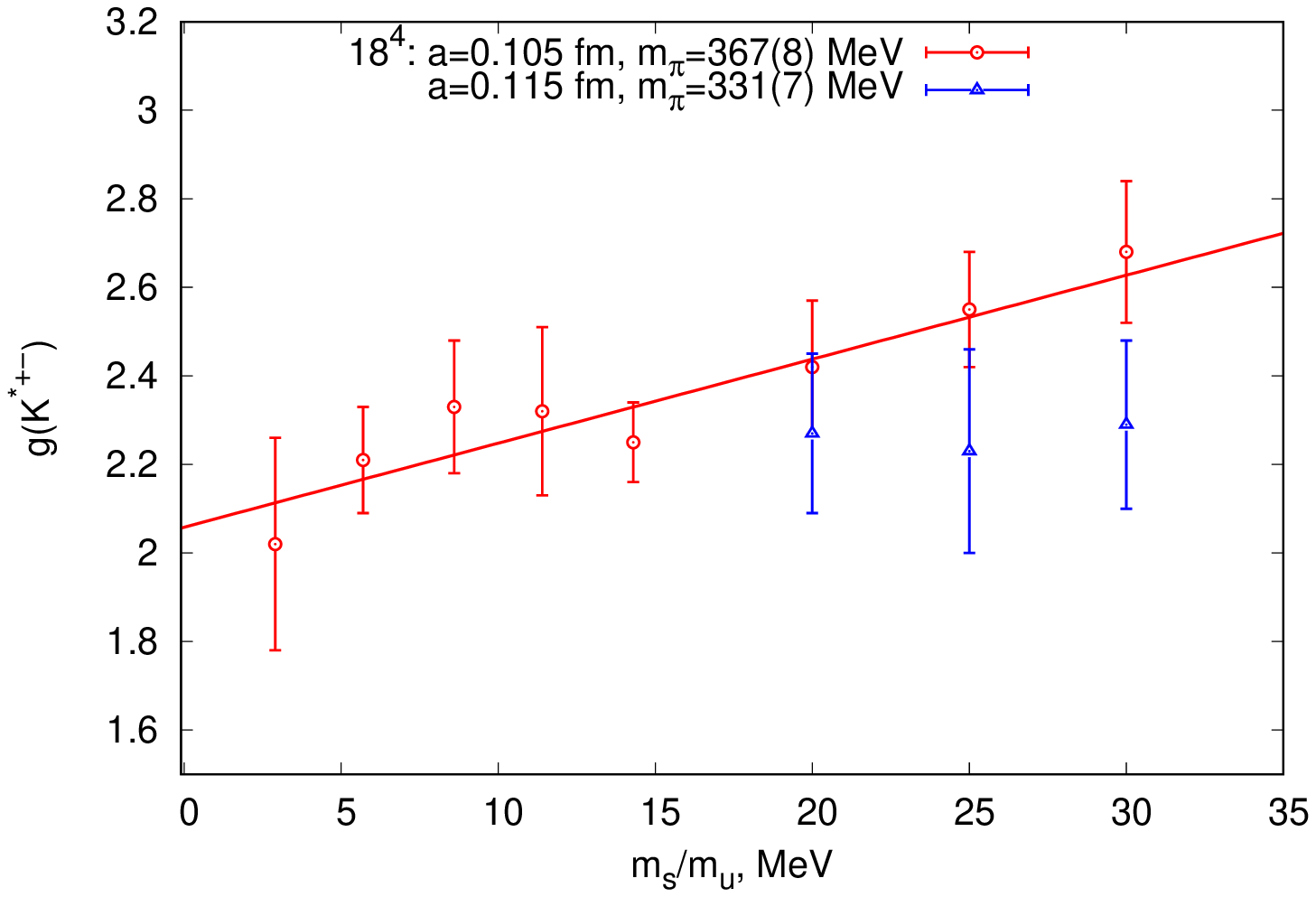}
\caption{The g-factor of the  vector $K^{*\pm}$ meson   obtained in this calculations for the lattice spacing $a=0.105\ \fm$ and pion mass $m_{\pi}=367(8)\ \MeV$ in comparison with the result of our previous work \cite{Luschevskaya:2017} for the lattice spacing $a=0.115\ \fm$ and   pion mass $m_{\pi}=331(7)\ \MeV$.}
\label{Fig:g_from_ms_mu}
\end{center}
\end{figure}
  \begin{table}
\begin{ruledtabular}
\begin{tabular}{cccccc}
  $m_s/m_u$  &  $g$-factor  &      $\chi^2$/n.d.f. &  n.d.f.    & p-value        & $eB(\GeV^2)$\\
\hline
    $2.9$      & $2.02 \pm 0.24$      & $1.321$  &   $1$   &   $0.251$    & $[0:0.41]$\\
   $5.7$     &  $2.21 \pm 0.12 $      & $0.633$   &   $1$  &    $0.426$   & $[0:0.41]$\\
   $8.6$     &  $ 2.33\pm 0.15 $      & $0.821$  &  $1$     &    $0.365$   & $[0:0.41]$\\
   $11.4 $   &  $2.32 \pm 0.19$       & $2.199$ &   $1$   &   $0.138 $  & $[0:0.41]$\\
   $14.3 $   &  $2.25 \pm 0.09$     & $2.005 $   &   $2$   &    $0.135$   & $[0:0.62]$\\
   $20$  &  $2.42\pm 0.15$          & $2.424$  &   $2$   &   $0.089$    & $[0:0.62]$\\
   $25$  &  $2.55\pm 0.13$          & $1.639$  &   $2$   &   $0.194$    & $[0:0.62]$\\
   $30$  &  $2.68 \pm 0.16$         & $1.678 $    & $2$   &   $0.187$    & $[0:0.62]$\\
\end{tabular}
\caption{The g-factor  of the vector $K^{*\pm}$ meson   depending on the $m_s/m_u$ value for the lattice volume $18^4$, lattice spacing $a=0.105\ \fm$ and the pion mass $m_{\pi} = 367\pm 8 \ \MeV$. The range of the magnetic field used for the fitting procedure is shown in the last column. The parameters of the fits are also represented in column three to five.}
\label{Table:gfactor:Kg1_S-1}
\end{ruledtabular}
\end{table}
The values of g-factor obtained from the fits are collected  in Table \ref{Table:gfactor:Kg1_S-1} and represented in Fig.~\ref{Fig:g_from_ms_mu} for the lattice volume $18^4$, lattice spacing $a=0.105\ \fm$ and the pion mass $m_{\pi} = 367\pm 8 \ \MeV$.  There is an indication that the magnetic moment increases with the growth of $m_s/m_d$ ratio, but this result  preliminary and demands the increase of statistics,  investigations of lattice volume and lattice spacing effects. In Fig.~\ref{Fig:g_from_ms_mu} we also show our previous results obtained for the coarser lattice spacing  $a=0.115\ \fm$ and  smaller pion mass  $m_{\pi}=331\pm 7\ \MeV$.

Our value g-factor obtained for   $m_s/m_d=20$ agrees   with     the other lattice calculations based on 2-point correlation functions \cite{Luschevskaya:2017,Lee:2008} and the prediction of QCD sum rules $2.0\pm 0.4$ \cite{Aliev:2009}. It is a little bit higher than the value $g=2.23$ which was found from 3-point correlation functions on the lattice \cite{Hedditch:2007}.

\section{Conclusions}

\label{subsec-6}

 \ \ \ \ We observe that for the case of the vector $K^{*0}$ meson  ($m_s/m_d=20,30$) the energy of the spin component $S_z=+1$ increase with the magnetic field and ecreasing of the $m_s/m_d$ does not change this behavior.
 For the   $S_z=-1$ a slow decrease at small fields $eB\lesssim 0.3\, \GeV^2$ is observed because the magnetic moment contributes to the energy squared. With the diminishing of the $m_s/m_d$ ratio the absolute value of the  magnetic moment decreases and at
 $m_s/m_d\lesssim 5.7$ the energy behavior is determined by the magnetic dipole polarizability which has a negative value.

 We obtain that the magnetic moment of the $K^{*0}$ meson  is negative and decreases with the growth of the $m_s/m_d$ ratio.
For the case of $\bar{K}^{*0}$ meson the energy branches $S_z=+1$ and $S_z=-1$ change places in comparison with the energy branches of $K^{*0}$ meson, which means that the magnetic moment of $\bar{K}^{*0}$ meson is a positive quantity.
The value of the g-factor $g(K^{*0})$  in a quantitative agreement with other results \cite{ Hedditch:2007,Simonov:2013b}. 
We leave the extrapolations over pion mass, lattice volume and lattice spacing extrapolations   for further investigations.
The nonzero magnetic moment of the neutral vector $\bar{K}^*$ and $K^*$ mesons can be explained by
the asymmetry of the motion of strongly bound heavy and light quarks inside the meson that is exposed to the  magnetic field  of QCD scale.

In  the case of zero spin $S_z=0$ a decrease in energy is observed,  but we do not see that the energy for this spin component goes to zero.
For the $S_z = 0$ spin projection the magnetic dipole polarizability    remains positive and constant within errors for  all the values $m_s/m_d$ considered, which leads to a decrease in energy. We found  the values of the magnetic polarizabilities equal to $\beta_m = 0.187\pm 0.017\ \GeV^{-3}$ at $m_s/m_d=20$  and $\beta_m = 0.167\pm 0.019\ \GeV^{-3}$ at $m_s/m_d=30$. However, for zero spin  the magnetic  hyperpolarizabilities   begin to contribute at low fields. The magnetic   hyperpolarizability of the first order  has the opposite  effect on the  behavior of the meson energy,  because it is negative in magnitude, for example, at $m_s/m_d=30$ we obtain $\beta_m^{h1} = -0.198 \pm 0.049\ \GeV^{-5}$. In   general, there is an alternation of signs of the terms in formula \eqref{eq:K0:s-0:B8}. Therefore, the energy should not go   to zero as in the case of neutral vector $\rho$ mesons \cite{Luschevskaya:2017}.

We also investigated how the  magnetic dipole polarizability for $S_z=\pm 1$  changes depending on the $m_s/m_d$ ratio.
For $S_z=-1$ spin projection  it stays constant within the error bars. For the $m_s/m_d=20$   the dipole magnetic polarizability is equal to $-0.027\pm 0.003\ \GeV^{-3}$, for  $m_s/m_d=30$  we obtain $\beta_m = -0.024\pm 0.003\ \GeV^{-3}$.
For the $S_z=+1$ energy branch     the
   magnetic dipole polarizability  increases from $\beta_m = -0.026\pm 0.006\ \GeV^{-3}$ at $m_s/m_d=1$ till zero value $ \beta_m = -0.0002 \pm 0.007\ \GeV^{-3}$ at $m_s/m_d=30$.

The different values of the magnetic polarizability for the $S_z=-1$ and $S_z=+1$   spin projections at  large    $m_s/m_d$ values  may be the lattice volume or lattice spacing effect. The $\beta_m$ for the $S_z=-1$ and $S_z=+1$   spins has to be equal due to parity concervation at the same value of $m_s/m_d$ ratio, but their difference does not influence our qualitative results on the lepton asymmetry or energy behavior.

We calculate the tensor polarizability for the case of $K^{*0}$ meson. The negative values of $\beta_t$ demonstrate that the longitudinal polarization dominates and  the dileptons from  $K^{*0}$ and $\bar{K}^{*0}$ decay in their rest frame mainly    in the directions perpendicular to the magnetic field. Also we found that the absolute value of the tensor polarizability diminishes  with the increase of the heavy mass quark. Despite the errors are large, this may be an indication that the heavier the strange quark, the smaller the asymmetry of the emitted dileptons.

For the charged vector mesons $K^{*\pm}$ we explore the dependence of $g$ - factor from the   strange quark mass. There is some indication that $g$-factor value increases with the growth of $m_s/m_d$ value, but this result demands an increase of statistics and  investigation of the lattice spacing and lattice volume effects  which we leave as the goal for the future work. We obtain $g=2.4 \pm 0.15$ at $m_s/m_d=20$ for the  lattice spacing $a=0.105\ \fm$, this value agrees  with our previous result \cite{Luschevskaya:2017}   for the lattice spacing $a=0.115\ \fm$.
Also we do not observe the tachyon  mode,  for the all $m_s/m_d$ values the energy for the $qS_z=-1$ decreases and   shows the plato at large fields.

\begin{acknowledgments}
The authors are grateful to  FAIR-ITEP supercomputer center where these numerical calculations were performed.
We are thankfull to M.N.~Chernodub and V.A.~Goy for usefull discussions. 
\end{acknowledgments}

\bibliography{aapmsamp}

\appendix

\section{Parameters of the fits}

  \label{app1}

   \begin{table*}
\begin{ruledtabular} 
\begin{tabular}{ccccccc}
  $m_s/m_d$  &  $g$-factor & $\beta_m(\GeV^{-3})$  & $\chi^2$/n.d.f.& n.d.f. & p-value     & $eB\, (\GeV^2)$\\
\hline
    $1$     & -                    & $-0.084 \pm 0.055$     & $0.676$      & $1$   &   $0.411$     & $[0:0.41]$\\
    $1$      & -                   & $-0.019 \pm 0.035$     & $1.100$      & $2$   &   $0.333$     & $[0:0.62]$\\
    $1$      & -                   & $-0.026 \pm 0.006$     & $0.751$      & $3$   &   $0.522$     & $[0:0.82]$\\
    $1$      & -                    & $-0.032 \pm 0.007$    & $1.216$      & $4$   &   $0.302$     & $[0:1.03]$\\
    $1$     & -                     & $-0.031 \pm 0.006$    & $0.997$      & $5$   &   $0.418$     & $[0:1.23]$\\
\hline
   $5.7$    & -                     &  $-0.033 \pm 0.062 $  & $2.750$     & $1$   &    $0.097$    & $[0:0.41]$\\
   $5.7$     & -                    &  $-0.038 \pm 0.006 $  & $1.384$     & $2$   &    $0.251$    & $[0:0.62]$\\
   $5.7$     & -                    &  $-0.029 \pm 0.006 $  & $1.215$     & $3$   &    $0.302$    & $[0:0.82]$\\
   $5.7$     & -                    &  $-0.033 \pm 0.003 $  & $1.042$     & $4$   &    $0.384$    & $[0:1.03]$\\
   $5.7$     & $-0.119 \pm 0.208$   &  $-0.040 \pm 0.013$   & $1.254$     & $3$   &    $0.289$    & $[0:1.03]$\\
\hline
   $8.6$     &  $-0.057 \pm 0.192 $  & -                    & $0.821$     & $1$  &    $0.365$      & $[0:0.41]$\\
   $8.6$     &  $0.007 \pm 0.096 $  & -                     & $0.490$     & $2$  &    $0.613$      & $[0:0.62]$\\
   $8.6$     &  $-0.404\pm 0.149 $  & $-0.045\pm 0.010$     & $0.212$     & $2$  &    $0.809$      & $[0:0.82]$\\
   $8.6$     &  $-0.173\pm 0.175 $  & $-0.028\pm 0.011$     & $0.527$     & $3$  &    $0.664$      & $[0:1.03]$\\
   $8.6$     &  $-0.044\pm 0.106 $  & $-0.019\pm 0.005$     & $0.510$     & $4$  &    $0.728$      & $[0:1.23]$\\
\hline
   $11.4 $   &  $-0.244 \pm 0.194$  & -                     & $1.404 $     &    $1 $    & $0.236 $ & $[0:0.41]$\\
   $11.4 $   &  $-0.143 \pm 0.112$  & -                     & $1.049 $     &    $2 $    & $0.350 $ & $[0:0.62]$\\
   $11.4 $   &  $-0.008 \pm 0.005$  & -                     & $1.903 $     &    $3 $    & $0.127 $ & $[0:0.82]$\\
   $11.4 $   &  $-0.555\pm 0.118$  & $-0.047 \pm 0.008$     & $0.236$      &    $2$    & $0.790 $   & $[0:0.82]$\\
   $11.4 $   &  $-0.373\pm 0.154$  & $-0.032 \pm 0.010$     & $0.651$      &    $3$    & $0.582 $   & $[0:1.03]$\\
  $11.4 $    &  $-0.265\pm 0.100$  & $-0.024 \pm 0.005$     & $0.628$      &    $4$    & $0.642 $   & $[0:1.23]$\\
\hline
   $14.3 $   &  $-0.325 \pm 0.203$  & -                     & $1.464 $     &  $1$    & $0.226$ & $[0:0.41]$\\
   $14.3 $   &  $-0.214 \pm 0.118$  & -                     & $1.119 $     &  $2$    & $0.326$ & $[0:0.62]$\\
   $14.3 $   &  $ 0.008 \pm 0.089$  & -                     & $2.529$      &  $3$    & $0.055$ & $[0:0.82]$\\
   $14.3 $   &  $-0.634 \pm 0.120$  & $-0.045 \pm 0.008$    & $0.237$      &  $2$    & $0.789$ & $[0:0.82]$\\
   $14.3 $   &  $-0.454 \pm 0.152$  & $-0.031 \pm 0.009$    & $0.615$      &  $3$    & $0.605$ & $[0:1.03]$\\
   $14.3 $   &  $-0.378 \pm 0.097$  & $-0.025 \pm 0.004$    & $0.535 $     &  $4$    & $0.710$ & $[0:1.23]$\\
\hline
   $20$  &  $-0.442 \pm 0.157 $  &     -                     & $0.505$        & $1$  &   $0.477$   & $[0:0.41]$\\  
   $20$  &  $-0.347 \pm 0.097 $  &     -                     & $0.422$        & $2$  &   $0.656$   & $[0:0.62]$\\
   $20$  &  $-0.171 \pm 0.098 $  &     -                     & $1.059 $       & $3$  &   $0.365$   & $[0:0.82]$\\
   $20$  &  $-0.751 \pm 0.095 $  &    $-0.039\pm 0.006$      & $0.075 $       & $2$  &   $0.928$   & $[0:0.82]$\\
   $20$  &  $-0.608\pm 0.116$    &    $-0.028  \pm 0.006$    & $0.189$        & $3$  &   $0.904$   & $[0:1.03]$\\
   $20$  &  $-0.599\pm 0.076$    &    $-0.027  \pm 0.003$    & $0.142$        & $4$  &   $0.966$   & $[0:1.23]$\\
\hline
   $30$  &  $-0.639 \pm 0.241$   &     -                      & $1.052 $      & $1$  &   $0.305$   & $[0:0.41]$\\
   $30$  &  $-0.535 \pm 0.131$   &     -                      & $0.700 $      & $2$  &   $0.497$   & $[0:0.62]$\\
   $30$  &  $-0.336 \pm 0.106$   &    -                       & $1.300 $      & $3$  &   $0.273$   & $[0:0.82]$\\
   $30$  &  $-0.989 \pm 0.165$   &    $-0.344 \pm 0.008$      & $0.207 $      & $2$  &   $0.813$   & $[0:0.82]$\\
   $30$  &  $-0.814 \pm 0.158$   &    $-0.024 \pm 0.007$      & $0.325 $      & $3$  &   $0.807$   & $[0:1.03]$\\
   $30$  &  $-0.816 \pm 0.103$   &    $-0.024 \pm 0.003$      & $0.244 $      & $4$  &   $0.913$   & $[0:1.23]$\\
\end{tabular}
\caption{The value of $g$-factor and  the magnetic dipole polarizability $\beta_m$  of the $K^{*0}$ meson  for the spin projection $S_z=-1$  at different values of the  $m_s/m_d$ ratio with other fit parameters depending on the range of magnetic fields used for the fit.  
At $m_s/m_d\leqslant5.7$ for  the fits   performed by the formula  \eqref{eq:K0:s1:B2}   the magnetic dipole polarizability $\beta_m$ is shown only. For the fits performed  by formula \eqref{eq:K0:s1:B1} the   $g$-factor is represented, this 2-parametric fit was used at $eB\leqslant0.82\,\GeV^2$. At larger field ranges and $m_s/m_d$ ratios the 3-parametric fit by the formula \eqref{eq:pt3:K0:s1:B2} was utilized.
The number of degrees of freedom n.d.f. and the p-value of significance level   are shown in the fifth and  sixth  columns respectively. The range of field used for the fitting procedure is represented in the last column.
}
\label{Table:beta:K0_S-1_comparison}
\end{ruledtabular}
\end{table*}

 \begin{table*}
\begin{ruledtabular} 
\begin{tabular}{cccccccc}
  $m_s/m_d$  &  $\beta_m (\GeV^{-3})$  & $\beta_m^{h1}(\GeV^{-5})$ & $\beta_m^{h2}(\GeV^{-8})$ & $\chi^2$/n.d.f & n.d.f. & p-value   & $eB\, (\GeV^2)$\\
\hline
   $1$     & $0.206 \pm 0.108$    & $-0.206 \pm 0.260$   & -                  & $1.509$  &   $1$  &   $0.219$    & $[0:0.62]$\\
   $1$     & $0.185 \pm 0.022$    & $-0.155 \pm 0.030$   & -                  & $0.784$  &   $2$  &   $0.456$    & $[0:0.82]$\\
   $1$     & $0.127 \pm 0.021$    & $-0.069 \pm 0.018$   & -                  & $2.689$  &   $3$  &   $0.045$    & $[0:1.03]$\\
\hline
   $2.9$   & $0.170 \pm 0.076$    & -                    & -                  & $5.044$  &   $1$  &    $0.025$   & $[0:0.41]$\\ 
   $2.9$   & $0.121 \pm 0.019$    & -                    & -                  & $3.578$  &   $2$  &    $0.028$   & $[0:0.62]$\\ 
   $2.9$   & $0.188 \pm 0.032$    & $-0.162 \pm 0.044$   & -                  & $2.346$  &   $2$  &    $0.096$   & $[0:0.82]$\\
\hline
   $5.7$   & $0.172 \pm 0.055 $   &  -                   & -                  & $9.913$  &   $1$  &    $0.002$  &  $[0:0.41]$\\
   $5.7$   & $0.132 \pm 0.028 $   &  -                   & -                  & $8.446$  &   $2$  &    $0.0002$  &  $[0:0.62]$\\
   $5.7$   & $0.199 \pm 0.036 $   &  $-0.174 \pm 0.049$    & -                 & $4.406$  &   $2$  &    $0.012$  &  $[0:0.82]$\\
   $5.7$   & $0.232 \pm 0.044 $   &  $-0.328 \pm 0.100$   & $0.158 \pm 0.058$ & $3.759$  &  $2$   &    $0.023$   &  $[0:1.03]$\\
\hline
   $8.6$    &  $0.173 \pm 0.028$  &  - & - & $2.058 $    &  $1$   & $0.151 $ & $[0:0.41]$\\
   $8.6$    &  $0.114 \pm 0.017$  &  - & - & $5.899 $    &  $2$   & $0.003 $ & $[0:0.62]$\\
   $8.6$    &  $0.212 \pm 0.019$  & $-0.278 \pm 0.041$ & $0.127 \pm 0.024$ & $0.861 $    &  $2$   & $0.423 $ & $[0:1.03]$\\
\hline
   $11.4$   & $0.180 \pm 0.030  $  & - & -       &    $3.366$           & $1$    & $0.067$ & $[0:0.41]$\\
   $11.4$   & $0.230 \pm 0.039  $  & $-0.277 \pm 0.098$  & -       &    $1.998$           & $1$    & $0.157$ & $[0:0.62]$\\
   $11.4$   & $0.182 \pm 0.020  $  & $-0.148 \pm 0.027$  & -       &    $2.794$           & $2$    & $0.061$ & $[0:0.82]$\\
\hline
   $14.3$   &  $0.171 \pm 0.029  $  & -                 & -        &    $3.009$           & $1$    & $0.083$ & $[0:0.41]$\\
   $14.3$   &  $0.119\pm 0.016$   & -        & -          &    $6.754$                    & $2$    & $0.001$ & $[0:0.62]$\\ 
   $14.3$   &  $0.218 \pm 0.039$  & $-0.257 \pm 0.039$     & -          &    $1.805$       & $1$    & $0.179$ & $[0:0.62]$\\   
   $14.3$   &  $0.175 \pm 0.018$  & $-0.140 \pm 0.026$      & -          &    $2.198$      & $2$    & $0.111$ & $[0:0.82]$\\
   $14.3$   &  $0.237 \pm 0.046$  & $-0.406\pm 0.019$     & $0.263\pm0.184$  &   $1.463$  & $1$    & $0.226$ & $[0:0.82]$\\
   $14.3$   &  $0.206 \pm 0.019$  & $-0.167 \pm 0.047$  & $0.122 \pm 0.029$  &    $2.198$  & $2$ & $0.311$ & $[0:1.03]$\\
\hline
   $20$     & $0.157\pm 0.022$ &  & - & $1.238$                                          & $1$    & $0.266$    & $[0:0.41]$\\
   $20$    &  $0.110\pm 0.015$ & - & - & $3.699$                                         & $2$   & $0.025$    & $[0:0.62]$\\
   $20$    &  $0.196\pm 0.029$ & $-0.220\pm 0.073$    & -                   & $0.764$    & $1$    & $0.382$    & $[0:0.62]$\\
   $20$    &  $0.160\pm 0.016$ & $-0.123\pm 0.024$    & -                   & $1.086$    & $2$    & $0.338$    & $[0:0.82]$\\
   $20$    &  $0.211\pm0.035$  & $-0.343\pm0.141$     & $0.219\pm0.140$     & $0.631$    & $1$    & $0.427$    & $[0:0.82]$\\
   $20$    &  $0.187\pm 0.017$ & $-0.233\pm0.044$     & $ 0.107 \pm 0.028 $ & $0.530$    & $2$    & $0.588$    & $[0:1.03]$\\
\hline
   $30$    &  $0.138\pm0.028$  & -                    & -                   & $1.471$    & $1$    & $0.225 $   & $[0:0.41]$\\
   $30$    &  $0.103\pm0.013$  & -                    & -                   & $2.010$    & $2$    & $0.134 $   & $[0:0.62]$\\
   $30$    &  $0.075\pm0.011$  & -                    & -                   & $5.579$    & $3$    & $0.0008 $  & $[0:0.82]$\\
  $30$     & $0.145 \pm 0.016$ & $-0.107 \pm 0.023$   & -                   & $0.756$    & $2$    &  $0.450$   & $[0:0.82]$\\
  $30$     & $0.117 \pm 0.014$ & $-0.060 \pm 0.014$   & -                   & $1.700$    & $3$    &  $0.164$   & $[0:1.03]$\\
  $30$     & $0.098\pm0.011$   & $-0.038\pm 0.007$    &  -                  & $2.494$    & $4$    &  $0.041 $  & $[0:1.23]$\\
  $30$     & $0.167\pm0.019$   & $-0.198 \pm 0.049$   &  $0.089 \pm 0.031$  & $0.501$    & $2$    &  $0.606 $  & $[0:1.03]$\\
\end{tabular}
\caption{The magnetic dipole polarizability $\beta_m$, the magnetic hyperpolarizabilities   $\beta_m^{h1}$ and  $\beta_m^{h2}$ of the $K^{*0}$ meson  for the spin projection $S_z=0$  at different values of the  $m_s/m_d$ ratio with other fit parameters depending on the range of magnetic fields used for the fit.  
For the fits   performed by the formula  \eqref{eq:K0:s-0:B4} the magnetic dipole polarizability $\beta_m$ and  the magnetic hyperpolarizability of the first order  $\beta_m^{h1}$  are represented. For the fits performed  by the formula \eqref{eq:K0:s-0:B6} the magnetic  hyperpolarizability of the second order  $\beta_m^{h2}$ is also shown. The number of degrees of freedom n.d.f. and the p-value of significance level   are shown in the   sixth and seventh columns respectively. The range of field used for the fitting procedure is represented in the last column.
}
\label{Table:beta:K0_S0_comparison}
\end{ruledtabular}
\end{table*}

\begin{table*}
\begin{ruledtabular} 
\begin{tabular}{ccccccc}
  $m_s/m_d$  &  $g$-factor  & $\beta_m(\GeV^{-3})$  & $\chi^2$/n.d.f. & n.d.f. & p-value    & $eB\, (\GeV^2)$\\
\hline
    $1$      & -                  & $-0.084 \pm 0.055$       & $0.676$      & $1$   &   $0.411$     & $[0:0.41]$\\
    $1$      & -                   & $-0.019 \pm 0.035$      & $1.100$      & $2$   &   $0.333$     & $[0:0.62]$\\
    $1$      & -                   & $-0.026 \pm 0.006$      & $0.751$      & $3$   &   $0.522$     & $[0:0.82]$\\
    $1$      & -                   & $-0.032 \pm 0.007$      & $1.216$      & $4$   &   $0.302$     & $[0:1.03]$\\
    $1$      & -                   & $-0.031 \pm 0.006$      & $0.997$      & $5$   &   $0.418$     & $[0:1.23]$\\
\hline 
    $2.9$    & -                   & $-0.062 \pm 0.043$      & $0.589 $     & $1$   &   $0.443$   & $[0:0.41]$\\  
    $2.9$    & -                   & $-0.050 \pm 0.019$      & $0.331 $     & $2$   &   $0.718$   & $[0:0.62]$\\ 
    $2.9$    & -                   & $-0.036 \pm 0.003$       & $0.284 $     & $3$   &   $0.837$   & $[0:0.82]$\\
    $2.9$    & -                   & $-0.036 \pm 0.002$       & $0.216 $     & $4$   &   $0.930$   & $[0:1.03]$\\
\hline
   $5.7$     &  -                   & $-0.076 \pm 0.040$    & $1.673$      & $1$   & $0.196$     & $[0:0.41]$\\
   $5.7$     &  $-0.227\pm 0.225$   & -                      & $2.658$      & $2 $   & $0.070 $    & $[0:0.62]$\\ 
   $5.7$     &  -                  & $-0.027 \pm 0.028$       & $2.676$      & $2 $   & $0.069$    & $[0:0.62]$\\  
   $5.7$     &  -                  & $-0.039 \pm 0.006$       & $1.952$      & $3$   & $0.119$     & $[0:0.82]$\\
   $5.7$     &  $-0.191\pm 0.337$   & $-0.022 \pm 0.023$    & $2.723$       & $4$ &  $0.028$      & $[0:1.23]$\\
\hline
   $8.6$  &  $ -0.498\pm 0.372 $     & -                      & $3.262$      & $1$ &  $0.071$      & $[0:0.41]$\\
   $8.6$  &  $ -0.607\pm 0.072 $   & -                       & $1.204$      & $3$ &  $0.307$      & $[0:0.82]$\\
    $8.6$  &  $-0.337\pm 0.273$     & $-0.020 \pm 0.018$      & $1.205$      & $3$ &  $0.306$      & $[0:1.03]$\\
    $8.6$  &  $-0.367 \pm 0.198$   & $-0.018 \pm 0.012$      & $0.916$      & $4$ &  $0.453$      & $[0:1.23]$\\
\hline
   $11.4$ &  $-0.452\pm 0.285$   & -                         & $3.514$      &    $1$    & $0.061$ & $[0:0.41]$\\
   $11.4$ &  $-0.568\pm 0.134$   & -                         & $2.202$      &    $2$    & $0.111$ & $[0:0.62]$\\
   $11.4$ &  $-0.615\pm 0.065$   & -                         & $1.601$      &    $3$    & $0.187$ & $[0:0.82]$\\
    $11.4$ &  $-0.302\pm 0.201$   & $-0.019\pm 0.010$        & $1.062$      &    $3$    & $0.364$ & $[0:1.03]$\\
   $11.4$ &  $-0.357\pm 0.153$   & $-0.019\pm 0.010$        & $0.866$      &    $4$    & $0.484$ & $[0:1.23]$\\
\hline
   $14.3$ &  $-0.525 \pm 0.267$      & -                     & $3.076$      & $1$    & $0.079$ & $[0:0.41]$\\
   $14.3$ &  $-0.623 \pm  0.133$     & -                     & $1.882$      & $2$    & $0.152$ & $[0:0.62]$\\
   $14.3$ &  $-0.679 \pm  0.064$     & -                     & $1.418$      & $3$    & $0.235$ & $[0:0.82]$\\
   $14.3$ &  $-0.385 \pm 0.180$      & $-0.021\pm 0.011$    & $0.855$       & $3$    & $0.464$ & $[0:1.03]$\\
   $14.3$ &  $-0.449 \pm 0.143$      & $-0.016\pm 0.009$    & $0.741$       & $4$    & $0.564$ & $[0:1.23]$\\
\hline
   $20$   &  $-0.668 \pm 0.324$      & -                     & $2.764$      & $1$    & $0.096$    & $[0:0.41]$\\
   $20$   &  $-0.763 \pm 0.161$      & -                     & $1.593$      & $2 $   & $0.203 $   & $[0:0.62]$\\ 
   $20$   &  $-0.789 \pm 0.081$      & -                     & $1.083$      & $3$    & $0.355$    & $[0:0.82]$\\  
   $20$   &  $-0.580\pm 0.223$       & $-0.013 \pm 0.012$    & $848$        & $3$    & $0.467$    & $[0:1.03]$\\  
   $20$   &  $-0.606\pm 0.173$       & $-0.011 \pm 0.009$    & $0.650$      & $4$    & $0.627$    & $[0:1.23]$\\ 
\hline
  $30$  & $-0.919 \pm 0.294$         & -                     & $1.775$      & $1$    &  $0.183$   & $[0:0.41]$\\
  $30$  & $-0.952\pm 0.141$          & -                     & $0.908$      & $2$    &  $0.403$   & $[0:0.62]$\\
  $30$  & $-1.010 \pm 0.070$         & -                     &  $0.681$     & $3$    &  $0.563$   & $[0:0.82]$\\
  $30$  & $-0.914\pm 0.221$          & $0.004 \pm 0.010$     &  $0.679$     & $3$    &  $0.565$   & $[0:1.03]$\\
  $30$  & $-0.981 \pm 0.172$         & $0.0002 \pm 0.007$    &  $0.565$     & $4$    &  $0.688$   & $[0:1.23]$\\
\end{tabular}
\caption{
The value of $g$-factor and  the magnetic dipole polarizability $\beta_m$  of the $K^{*0}$ meson  for the spin projection $S_z=+1$  at different values of the  $m_s/m_d$ ratio with other fit parameters are represented. As for the case of $S_z=-1$ spin projection the 2-parametric fits by formulas \eqref{eq:K0:s1:B2} and \eqref{eq:K0:s1:B1} and 3-parametric fit by formula \eqref{eq:pt3:K0:s1:B2} were utilized, see the description  in  Table \ref{Table:beta:K0_S-1_comparison}.}
\label{Table:beta:K0_S+1_comparison}
\end{ruledtabular}
\end{table*}

\end{document}